\documentclass[aps,pre,twocolumn]{revtex4}
\usepackage{graphics}
\usepackage{graphicx}

\begin{document}


\title{Turbulence spreading and anomalous diffusion on combs}




\author{Alexander V. Milovanov}
\affiliation{ENEA National Laboratory, Centro Ricerche Frascati, I-00044 Frascati, Rome, Italy}
\affiliation{Max Planck Institute for the Physics of Complex Systems, 01187 Dresden, Germany}
	 
\author{Alexander Iomin}
\affiliation{Solid State Institute, Technion, Haifa, 32000, Israel}	
\affiliation{Max Planck Institute for the Physics of Complex Systems, 01187 Dresden, Germany} 

\author{Jens Juul Rasmussen}
\affiliation{Physics Department, Technical University of Denmark, DK-2800 Kgs.~Lyngby, Denmark}



\begin{abstract} 
This paper presents a simple model for such processes as chaos spreading or turbulence spillover into stable regions. In this simple model the essential transport occurs via inelastic resonant interactions of waves on a lattice. The process is shown to result universally in a subdiffusive spreading of the wave field. The dispersion of this spreading process is found to depend exclusively on the type of the interaction process (three- or four-wave), but not on a particular instability behind. The asymptotic transport equations for field spreading are derived with the aid of a specific geometric construction in the form of a comb. The results can be summarized by stating that the asymptotic spreading pursues as a continuous-time random walk (CTRW) and corresponds to a kinetic description in terms of fractional-derivative equations. The fractional indexes pertaining to these equations are obtained exactly using the comb model. A special case of the above theory is a situation when two waves with oppositely directed wave vectors couple together to form a bound state with zero momentum. This situation is considered separately and associated with the self-organization of wave-like turbulence into banded flows or staircases. Overall, we find that turbulence spreading and staircasing could be described based on the same mathematical formalism, using the Hamiltonian of inelastic wave-wave interactions and a mapping procedure into the comb space. Theoretically, the comb approach is regarded as a substitute for a more common description based on quasilinear theory. Some implications of the present theory for the fusion plasma studies are discussed and a comparison with the available observational and numerical evidence is given.         
\end{abstract}

\maketitle

\section{Introduction}

Turbulence spreading \cite{Garbet} is the spillover into surrounding stable areas of turbulent motions excited at some location. The phenomenon characterizes both fluid and plasma turbulence and has been reported experimentally and/or numerically for a variety of systems and physical conditions, with a wealth of data spanning solar and astrophysics \cite{Tao,Brummel}, geophysics \cite{Large,Plume} and magnetic confinement fusion \cite{Itoh,Naulin,Zonca06,Lin}. In fusion grade plasmas, turbulence occurring in the linearly active (unstable) regions of a plasma can penetrate into the linearly inactive (stable) regions of the same plasma, where it can modify transport scalings \cite{Garbet,Itoh} and eventually deteriorate confinement \cite{Naulin,Zonca06,Lin,Gurk05,Gurk06,Gurk07,Guo,Migliano,Kwon15}. More recently, it has been discussed \cite{Korean,Nature,Singh} that turbulence spreading can mediate the global self-organization of L-mode tokamak plasma into a marginally stable state and that it underlies such phenomena as the rise and decay of transport barriers \cite{Singh,Yagi,Zeng,Zonal}, scrape-off layer (SOL)-core and SOL-edge coupling \cite{Short,Korean}, avalanche transport \cite{Politzer,Tokunaga,Canada,Hein}, and the staircase self-organization \cite{DF2010,DF2015,DF2017,Horn2017,Garbet21,Neg,Ashour,Fang,Choi,Van}. Indirect evidence of turbulence spreading may be obtained from, e.g., the breakdown of gyro-Bohm transport scaling \cite{Lin02,Korean}, the breakdown of Fick's law \cite{Korean,Rev15}, the broadening of SOL \cite{Chu,Wu}, and the transport shortfall problem \cite{Nature,Short,Singh}. Further evidence comes along with cold-pulse phenomenology \cite{Pulse} and the observation of internal rotation reversal \cite{Pulse,Hariri} and pulse-polarity reversal \cite{Hariri,Rice13,Rice}, yet among other observations \cite{Zonca06,Korean,Nature}. 

On the theory side, a brief account of the existing literature suggests the absence of a unifying theory of turbulence spreading that would apply in all cases. A paradigmatic approach to turbulence spreading (e.g., Refs. \cite{Garbet,Itoh,Naulin,Zonca06,Lin,Gurk05,Gurk06,Gurk07,Guo,Migliano,Singh,Hein,Korean}) relies on a conjecture that the transport of turbulence intensity can be described using a nonlinear diffusion-reaction equation with sources and sinks\textemdash akin to the Fisher-Kolmogorov-Petrovsky-Piskunov (F-KPP) equation \cite{Fisher,KPP,Canc}. By mastering a suitable nonlinearity in the diffusion coefficient and in respective driving and damping terms one succeeds based on this equation to reproduce a number of complex situations of interest to fusion tasks (e.g., Refs. \cite{Zonca06,Gurk07,Singh,Hein,Migliano}) under certain rather weak conditions. A criticism raised against this approach, however, is that it relies on a phenomenological characterization of turbulence spreading and uses a Gaussian propagator of turbulence intensity, which is not at all obvious in nonequilibrium systems. Other approaches \cite{Malkov,Malkov_NF,PoP13,Kosuga} advocate an idea that turbulence spreading can be understood as a transport problem for fluctuation pulses, suggesting a theoretical description in terms of Burgers' equation with noise. In the context of tokamak plasma, the Burgers' model can be formulated \cite{Korean,Kosuga} so that it includes both inward spreading from SOL into edge and outward spreading from edge into SOL. In the latter case it mimics the turbulence overshoot by coherent structures, vortexes and blob-filaments \cite{Zweben,Manz,Manz20,Ippolito}. In the outer core$-$inner edge region, the observed phenomenology \cite{Naulin,Hariri,Naulin07} of turbulence spreading is captured by the Hasegawa-Wakatani model \cite{HW1,HW2,Horton} of plasma edge turbulence, revealing a spatially anisotropic transport with bursts \cite{JJR,PLAN,Basu1,Basu2,PLA14}. 

In the realm of wave turbulence \cite{Sagdeev,QL}\textemdash composed of a large number of interacting waves with a distribution of frequencies and wave vectors\textemdash the problem of turbulence spreading can be reconciled \cite{PRE24} with the fundamental problem of quantum localization of dynamical chaos \cite{Sh93}. The latter problem has been the subject of very extensive theoretical treatments (e.g., Refs. \cite{PS,Flach,Skokos,Wang,Fishman,Iomin,Basko,EPL,Ivan,PRE14,QW,Many,Fistul}), in terms of the nonlinear Schr\"{o}dinger equation with random potential and the Hubbard model. The results from those investigations can be summarized by a critical value of the nonlinearity parameter, above which the nonlinear field spreads indefinitely along the lattice, and below which it is spatially localized similarly to the linear field. Authors of Refs. \cite{Iomin,EPL,QW} suggested that the dynamics of this spreading process could be characterized using continuous-time random walks (CTRWs)\textemdash leading to a theoretical description in terms of fractional-derivative equations \cite{Klafter,Sokolov,Rest}.      

By CTRW\textemdash introduced in physics by Montroll and Weiss \cite{CTRW1,CTRW2}\textemdash one means a random-walk process with a distribution of either step-sizes (Gaussian or L\'evy \cite{Gnedenko}) or waiting times between steps (Poisson or fat-tailed \cite{Klafter,Sokolov}) or both. As such, CTRWs underpin nondiffusive generalizations \cite{Bouchaud,Klafter,Zum,Andrey} of the Brownian random walk \cite{Bouchaud,Kampen}, enabling outstanding applications in such areas as disordered media \cite{Lax,Jons,Dyre,bAHa00}, dielectric relaxation \cite{Jons,Coffey,PRB07} and magnetic confinement fusion \cite{VM044,Carreras04,VM06,Carreras06}.  

In this study, we build upon these ideas and devise a theory of turbulence spreading based on the Hamiltonian of inelastic wave-wave interactions on a lattice. By applying Fermi's golden rule \cite{Golden} we relate the transport problem for turbulence intensity to a first-principal description of the nonlinear field. Following the analysis of Ref. \cite{PRE24}, we show that the asymptotic spreading law is determined exclusively by whether the interactions are three- or four-wave-like, regardless of the specific instability behind. Finally, we obtain asymptotic kinetic equations for turbulence spreading by setting the transport problem on a Dirac comb.     

In mathematics, a Dirac comb \cite{bAHa00} is a geometric representation of the pulse function 
\begin{equation}
{\mathcal{C}}_{\Lambda} (x): = \sum_{m = -\infty}^{+\infty} \delta (x - m\Lambda),  
\label{Pulse} 
\end{equation} 
where $\Lambda$ is the period of the comb (period between consecutive pulses), $m=0,\pm 1,\pm 2,\dots$ is an integer counter, and $\delta (x)$ denotes the Dirac delta-function.  

A Dirac comb (see Fig.~1) consists of a central backbone along the $x$ axis and infinite side branches (fingers or teeth) in the $y$ direction. Note that the number of side branches is countable (because the set of all integer numbers is countable). 

From a dynamical perspective, combs and their generalizations provide a convenient approximation to CTRWs (because side branches operate as dynamical traps with a distribution of exit times). On the other hand, because combs are loopless structures (similarly to Bethe lattices \cite{bAHa00}), their transport properties can in many situations be determined exactly \cite{WhBa84,GeGo85,WeHa86}.

\begin{figure}
\includegraphics[width=0.53\textwidth]{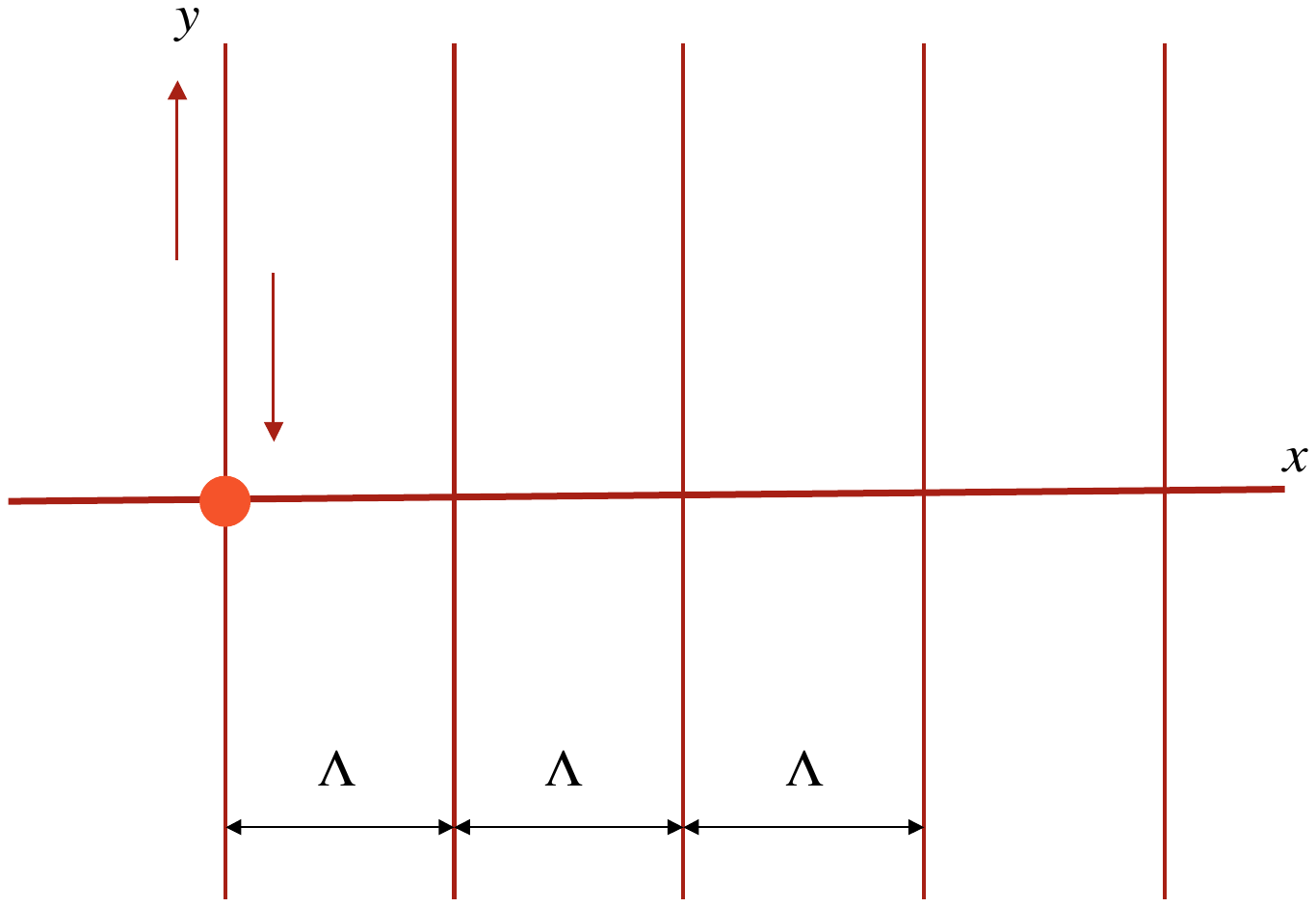}
\caption{\label{} The Dirac comb. The central horizontal line is the backbone, the vertical lines are the side branches\textemdash the so-called fingers or teeth of the comb. The bright red circle on the left is the origin. $\Lambda$ is the spacing between neighboring teeth, $x$ is the coordinate along the backbone, and $y$ is the coordinate in fingers.    
}
\end{figure}

The appreciation of combs in physics began with the work of Ziman \cite{Ziman}, who introduced them as a simplification of the percolation model of de Gennes \cite{Gennes}. In Ziman's description, a percolation cluster is thought of as composed of a conducting path, which corresponds to the backbone, and side branches, which represent the dead-ends of the cluster. Then at some level of idealization one draws a connected graph with infinite teeth that resembles a comb. Such comb-like graphs as in Fig.~1 have been applied in a basic theory of CTRWs as an alternative to fractal lattices (Refs. \cite{WeHa86,bAHa00,IoMeHo2018}). 

A remarkable feature about combs is that they capture much of the actually observed signatures of anomalous transport in disordered systems (e.g., Refs. \cite{bAHa00,So12,IoZa16}). In fusion grade plasmas, the use of combs is less noted, with only few exceptions, among which we specifically mention a model of the plasma staircase \cite{PRE18} and a demonstration, supported by numerical evidence, of weak localization of plasma avalanches \cite{PRE21}.  

The paper is organized as follows. The asymptotic scaling laws for turbulence spreading are derived first (Secs.~II\,A and II\,B), followed by a derivation of the exit-time distributions in Sec.~II\,C. These derivations suggest a special case of the zero-frequency resonance, which is considered separately in Sec.~III. The comb model is formulated in Sec.~IV. Section~V introduces the basic transport model and its 1D reduction, the partial cases of which are analyzed in Sec.~VI. Next, the fractional relaxation equation is discussed (Sec.~VII). We summarize our findings in Sec.~VIII. Some auxiliary results pertaining to the comb model are presented in Appendix A.        

\section{General case}

We envisage turbulence as a superposition of a large number of interacting waves with the dispersion relation $\omega_{{\bf k}_i} = \omega_i ({\bf k}_i)$, where $\omega_{{\bf k}_i}$ is the frequency of the $i$th wave, and ${\bf k}_i$ is the wave vector. The conservation of energy and momentum through the interaction process implies that the interaction cross-section has sharp peak whenever there is a resonance among the waves involved and vanishes otherwise. Respectively for three- and four-wave interactions the conditions for a resonance read \cite{Sagdeev}
\begin{equation}
\omega_{\bf k} = \omega_{{\bf k}_1} + \omega_{{\bf k}_2},\ \ \ {\bf k} = {\bf k}_1 + {\bf k}_2,
\label{R3} 
\end{equation} 
and 
\begin{equation}
\omega_{{\bf k}_1} + \omega_{{\bf k}_2} = \omega_{{\bf k}_3} + \omega_{{\bf k}_4},\ \ \ {\bf k}_1 + {\bf k}_2 = {\bf k}_3 + {\bf k}_4,
\label{R4a} 
\end{equation} 
\begin{equation}
\omega_{{\bf k}_1} = \omega_{{\bf k}_2} + \omega_{{\bf k}_3} + \omega_{{\bf k}_4},\ \ \ {\bf k}_1 = {\bf k}_2 + {\bf k}_3 + {\bf k}_4.
\label{R4b} 
\end{equation}
The triad equations~(\ref{R3}) [and likewise the quartic Eqs.~(\ref{R4a}) and~(\ref{R4b})] can be thought of as the defining equations for wave vectors ${\bf k}_i$. These equations may or may not have a solution as the dispersion relation $\omega_{{\bf k}_i} = \omega_i ({\bf k}_i)$ imposes a nontrivial constraint on the admissible frequencies and wave vectors. If a solution exists, it may be of three types. One type is associated with parametric decay instability, i.e., a process when strong falling wave decays into two or more lower-frequency waves such as in Figs.~2 and~3, top. This process is thresholded in that the falling wave's amplitude must exceed a certain critical value for the actual breakup to occur. Another type is exact opposite process of the above when two or more waves merge into one stronger wave. Finally, and this is specific to four-wave interactions, two falling waves may participate in an inelastic scattering process such as in Fig.~3, bottom, in which process the energy and momentum are merely redistributed among the participating waves, while the number of waves is preserved. These decay, merger and scattering processes are all important in far-from-equilibrium plasma systems, particularly in regimes underlying the expansion of turbulence into surrounding stable areas (e.g., Refs. \cite{Gurk06,Gurk07,PRE24}). 

In this paper's work, we are interested in the expansion of a small puff of turbulence excited locally from the exterior (for instance, by injecting a strong pump wave into a plasma \cite{QL}). The injected power is assumed to be sufficient to trigger a cascade of secondary decay events, if three- or four-wave, or both, thus populating the system with a certain initial number of unstable modes. Once locally excited, the instability can propagate to large spatial scales when nonlinearity, such as a spatial pressure inhomogeneity  \cite{Horton,JJR}, couples different waves, and energy and momentum conservation conditions are met at each step of the nonlinear interaction process.   

\begin{figure}
\includegraphics[width=0.54\textwidth]{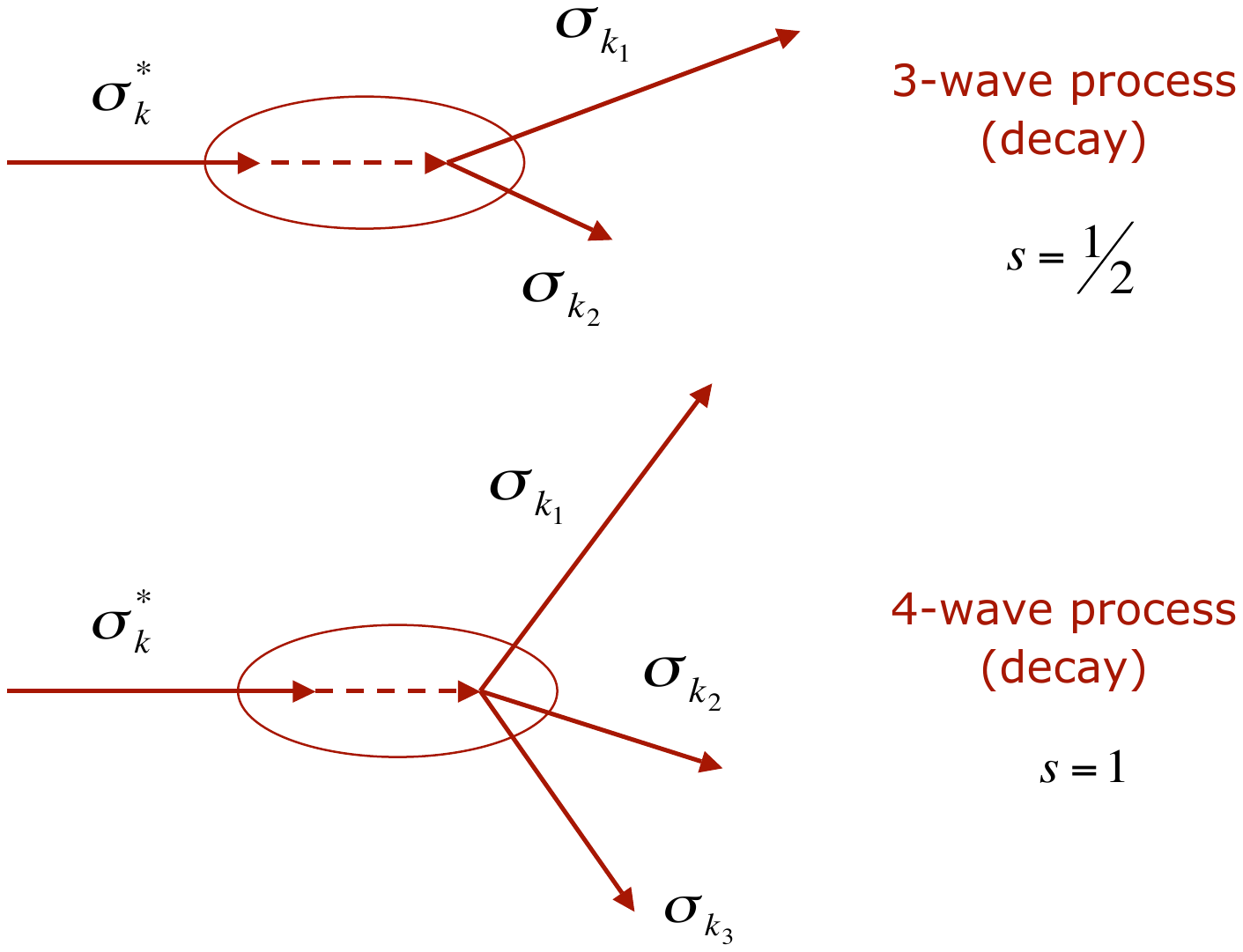}
\caption{\label{} Resonant decay processes. Top: A three-wave decay process when a falling wave breaks into two offspring waves. Bottom: The analogue four-wave process when a falling wave breaks into three offspring waves. $\sigma_k$ and other sigmas alike denote complex amplitudes of the various processes involved and are explained in the main text [see the paragraph after Eq.~(\ref{3W})]. The $s$ index pertains to Eqs.~(\ref{RR}) and~(\ref{WT}). The oval, with the dashed line inside, reminds that the interaction processes above refer to waves, not particles, meaning these processes might {\it not} be spatially localized.   
}
\end{figure}
\begin{figure}
\includegraphics[width=0.54\textwidth]{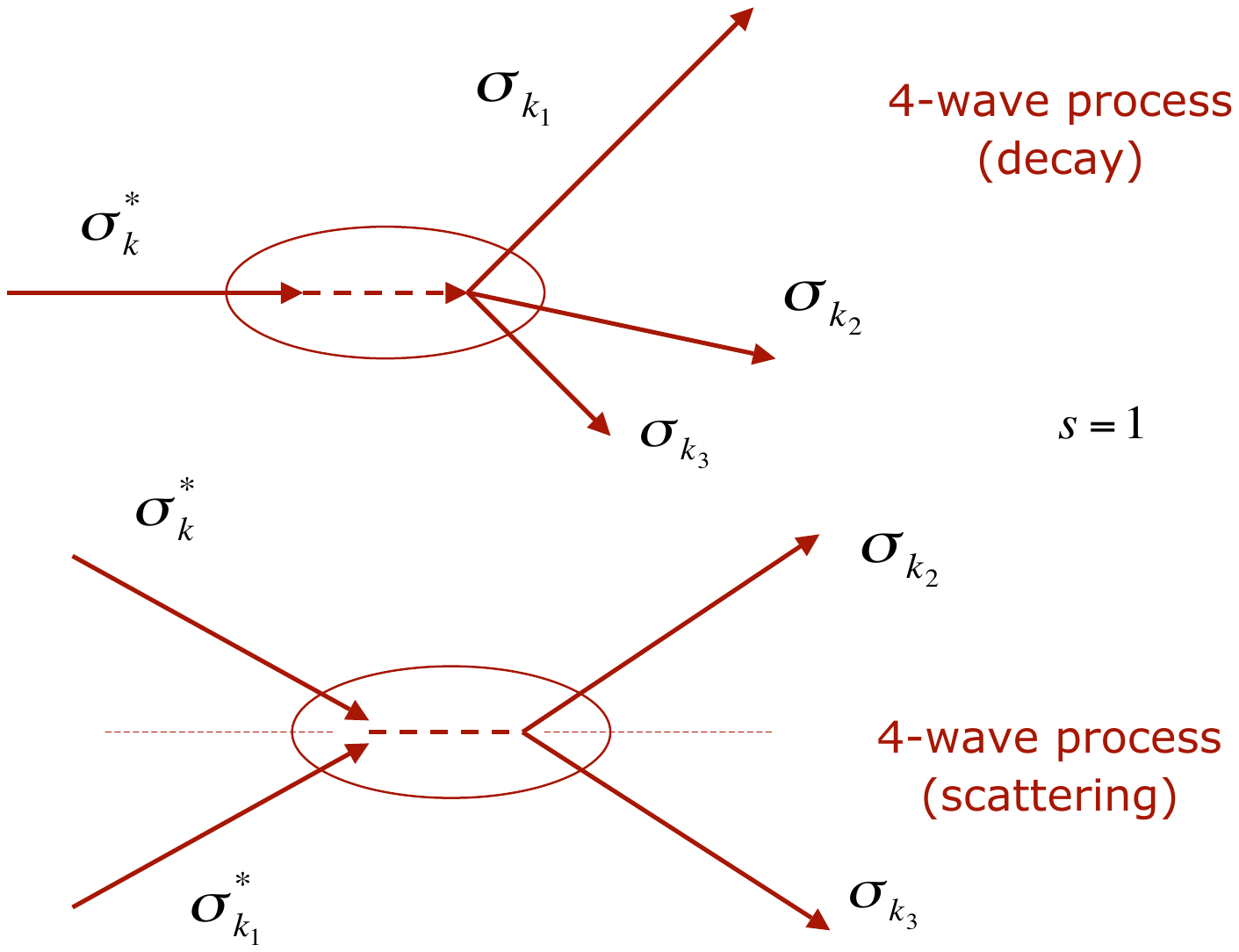}
\caption{\label{} A four-wave decay (top) versus scattering (bottom) process: a schematic illustration. Same notation as in Fig.~2 above.   
}
\end{figure}

In fusion applications, one encounters a situation according to which noninteracting waves can propagate freely in the preferred direction (in the case of drift waves it is the poloidal direction in a tokamak), while being linearly localized in the perpendicular (radial) direction. In tokamak geometry, a running wave must meet the known constraints \cite{Horton,Heid} imposed by poloidal and toroidal periodicities, through which periodicities the noninteracting waves are attracted to rational magnetic flux surfaces. When nonlinear interactions are allowed, then the main effect these interactions have on the wave field is to induce instability on neighboring flux surfaces, which is equivalent to an expansion of the wave field in the direction of radial localization. It is this process of instability expansion driven by nonlinear interactions which we associate with the phenomenon of turbulence spreading. The implication is that the spreading process (in its wave formulation) is fundamentally {\it anisotropic}\textemdash it occurs in the direction {perpendicular} to the wave vectors ${\bf k}_i$, while the interactions propagate {along} the ${\bf k}_i$'s. The situation is somewhat different from an (apparently similar) nonlinear Anderson problem \cite{PS,Flach,EPL}, where nonlinearity squeezes the wave field in the direction of the ${\bf k}$ vector.

In this study, we approximate the phenomena of turbulence spreading with a simple dynamical model as follows. It is assumed that all interactions occur on an infinite 1D lattice, which is aligned in the direction of wave propagation. The spreading is in perpendicular (radial) direction and comes about as a result of nonlinear interaction among the different unstable modes, provided the necessary conditions of energy and momentum conservation [Eqs.~(\ref{R3})$-$(\ref{R4b})] are satisfied. We argue it is the fundamental structure of these interactions that determines the resulting asymptotic transport scalings. 

With these implications in mind, we abandon hereafter the vector notation in Eqs.~(\ref{R3})$-$(\ref{R4b}), and continue with a scalar description instead. In particular, we use $k$ in place of $\bf k$, and we define this as the scalar product $k = {\bf k}\cdot \hat {\bf {e}}$, where $\hat {\bf {e}}$ is a unit vector along the lattice. Note that $k$ may have both signs depending on whether $\bf k$ looks along or against the $\hat {\bf {e}}$ vector. Yet so, we refer to $k$ as ``wave vector" for simplicity. The reflection symmetry ${\bf k} \longleftrightarrow -{\bf k}$ is assumed, i.e., the 1D lattice on which the interactions occur is isotropic. With this last condition, the dispersion relation is simplified to $\omega_{{k}_i} = \omega_i (|{k}_i|)$, implying $\omega_{{k}_i} = \omega_{-{k}_i}$.

\subsection{Three-wave interactions}

The Hamiltonian of three-wave interactions on a 1D discrete lattice reads (see, e.g., Refs. \cite{Sagdeev,Bers})
\begin{equation}
H = H_{0} + H_{\rm int}, \ \ \ H_0 = \frac{1}{2}\sum_k \omega_k \sigma^*_k \sigma_k,
\label{H0} 
\end{equation}
\begin{equation}
H_{\rm int} = \frac{1}{3}\sum_{k, k_1, k_2} V_{-k, k_1, k_2} \sigma^*_{k} \sigma_{k_1} \sigma_{k_2}\delta_{-k+k_1+k_2, 0},
\label{3W} 
\end{equation}
where $H_0$ is the Hamiltonian of noninteracting waves, $H_{\rm int}$ is the interaction Hamiltonian, $\sigma_k = \sigma_k (t)$ are complex amplitudes which represent a wave process with frequency $\omega_k$ and wave vector $k$ and which may depend on time $t$ in general, $\omega_{-k} = \omega_k$ thanks to $\omega_k = \omega (|{k}|)$, $\sigma_{-k} = \sigma_k^*$ by way of temporal translation symmetry \cite{Sagdeev}, the asterisk denotes complex conjugate, the set of all $k$'s is countable (in a discrete model), $V_{-k, k_1, k_2}$ are complex coefficients which characterize the cross-section of an interaction process with triad couplings $k = k_1 + k_2$, and $\delta_{-k+k_1+k_2, 0}$ is the Kronecker delta, which accounts for resonant character of these interactions. Remembering that the lattice is isotropic, i.e., $\omega_k = \omega (|{k}|)$ for all $k$, the summation in Eq.~(\ref{H0}) is performed over {both} positive and negative values of $k$ allowed by the dispersion relation. For the same reason, the summation in Eq.~(\ref{3W}) goes over all positive and negative values of $k$, $k_1$, $k_2$ allowed jointly by the dispersion relation $\omega_{k_i} = \omega_i (|{k_i}|)$ and the conservation laws. This, together with the sign-reversal conditions $\sigma_{-k} = \sigma_{k}^*$ for all $k$ and $V_{-k, k_1, k_2} = V^*_{k, -k_1, -k_2}$ for all $k$, $k_1$, and $k_2$, guarantees that the Hamiltonian in Eqs.~(\ref{H0}) and~(\ref{3W}) is real and well-defined. The cubic terms $\sigma^*_{k} \sigma_{k_1} \sigma_{k_2}$ in Eq.~(\ref{3W}) incorporate all admissible three-wave interaction processes, of both the decay type and the merger type, in which processes $\sigma^*_{k}$ represents the falling wave, and $\sigma_{k_1}$ and $\sigma_{k_2}$ represent the offspring waves (see Fig.~2, top). 

Given the Hamiltonian in Eqs.~(\ref{H0}) and~(\ref{3W}), one applies the canonical equations 
\begin{equation}
\dot{\sigma}_k = i\frac{\partial H}{\partial\sigma_k^*}, \ \ \  \dot{\sigma}_k^* = -i\frac{\partial H}{\partial\sigma_k}
\label{Can} 
\end{equation}
to obtain the equations of motion for the complex amplitudes ${\sigma_k}$, i.e., 
\begin{equation}
\dot{\sigma}_k = i\omega_k\sigma_k + iV^*\sigma_{k_1} \sigma_{k_2}.
\label{DEM} 
\end{equation}
The equations for $\sigma_{k_1}$ and $\sigma_{k_2}$ are obtained either directly from the Hamiltonian $H = H_0 + H_{\rm int}$ by applying the canonical equations, or by switching indexes in the equations of motion~(\ref{DEM}) on account of the resonance condition $k=k_1 + k_2$ and the general rule $\sigma_{-k} = \sigma_k^*$. The end result is 
\begin{equation}
\dot{\sigma}_{k_1} = i\omega_{k_1}\sigma_{k_1} - iV\sigma_{k} \sigma_{k_2}^*,
\label{DEM1} 
\end{equation}
\begin{equation}
\dot{\sigma}_{k_2} = i\omega_{k_2}\sigma_{k_2} - iV\sigma_{k} \sigma_{k_1}^*.
\label{DEM2} 
\end{equation}
In the above, $V \equiv V_{k, -k_1, -k_2}$ for simplicity, and the upper dot denotes time differentiation. 

If the field is spread over $\Delta n \gg 1$ states (in radial direction), then the conservation of the total probability 
\begin{equation}
\sum_{n=0}^{\Delta n} |\sigma_n|^2 \simeq \int_0^{\Delta n} |\sigma_n|^2 dn = 1
\label{TP} 
\end{equation}
implies that the density of the probability is small and is inversely proportional to $\Delta n$, i.e., $|\sigma_n|^2 \simeq 1/\Delta n$. 

It is understood that the evolution of the wave field in time is due to nonlinear coupling among the resonant modes. The rate of excitation of a newly involved mode is obtained as intensity of this coupling process, i.e., $R \simeq \rho |V^*|^2|\sigma_{n_1} \sigma_{n_2}|^2$, where $\rho$ is a coefficient. Using dynamical Eqs.~(\ref{DEM})$-$(\ref{DEM2}), one gets 
\begin{equation}
R \simeq \rho |V|^2|\sigma_n|^4\simeq \rho |V|^2/(\Delta n)^2,
\label{Rate} 
\end{equation}
where the scaling $|\sigma_n|^2 \simeq 1/\Delta n$ has been considered. In writing Eq.~(\ref{Rate}) we have tacitly assumed that the number of modes in Eqs.~(\ref{DEM})$-$(\ref{DEM2}) is large enough to render $\Delta n \gg 1$. The latter condition guarantees $R \ll \omega_k$ for the majority of modes, i.e., the radial expansion is slow compared to a typical frequency of the running wave. On the other hand, the resonant character of interactions dictates 
\begin{equation}
R = d\Delta n/ dt
\label{Rate0} 
\end{equation}
consistently with Fermi's golden rule \cite{Golden} for transitions between states. Combining Eqs.~(\ref{Rate}) and~(\ref{Rate0}), one obtains
\begin{equation}
d\Delta n / dt = \rho |V|^2/(\Delta n)^2. 
\label{Rate2} 
\end{equation}
Integrating over time in Eq.~(\ref{Rate2}), one is led to $(\Delta n)^3 = 3\rho |V|^2 t$, yielding
\begin{equation}
(\Delta n)^2 = (3\rho)^{2/3} |V|^{4/3} t^{2/3}.
\label{SpL} 
\end{equation}
One sees that the asymptotic spreading is subdiffusive: $(\Delta n)^2$ grows slower-than-linear with time as $t\rightarrow+\infty$. 

In the context of fusion plasma, one usually looks into a spatial spread $\Delta x$ instead of the number of states, $\Delta n$. However, if the linear field is spatially localized in radial direction, and if the localization mechanism is such as described by Garbet {\it et al.} \cite{Garbet} (i.e., stickiness to rational flux surfaces), then $\Delta n$ will be proportional to radial spread, i.e., $\Delta n \propto \Delta x$. From Eq.~(\ref{SpL}) one infers $\Delta x \propto t^{1/3}$. This is remarkable, as the latter scaling coincides with the scaling \cite{Itoh,Gurk05} deduced from the F-KPP equation.

\subsection{Four-wave interactions} 

In the case of four-wave interactions (see Fig.~3), the interaction Hamiltonian becomes 
\begin{equation}
H_{\rm int} = \frac{1}{4}\sum_{k, k_1, k_2, k_3} V_{-k, k_1, k_2, k_3} \sigma^*_{k}\sigma_{k_1}\sigma_{k_2}\sigma_{k_3}\delta_{-k+k_1+k_2+k_3, 0}
\label{4W} 
\end{equation}
and represents a next-order correction to the three-wave interaction Hamiltonian introduced in Sec. II\,A. As usual \cite{Sagdeev}, temporal translation symmetry dictates $\sigma_{-k} = \sigma_k^*$ for all $k$ and $V_{-k, k_1, k_2, k_3} = V^*_{k, -k_1, -k_2, -k_3}$ for all $k$, $k_1$, $k_2$, and $k_3$. It is assumed that the dispersion relation is symmetric with respect to the inversion $k_i \rightarrow -k_i$, i.e., $\omega_{{k}_i} = \omega_i (|{k}_i|)$, where ${k_i} = {\bf k}_i\cdot \hat {\bf {e}}$ on a 1D lattice. The summation in Eq.~(\ref{4W}) is performed over {all} admissible values of $k$, $k_1$, $k_2$, $k_3$ allowed by the dispersion relation and the conservation laws, that is any $k_i$ is included on an equal footing with $-k_i$, where $k_i = k, k_1, k_2$, $k_3$. This guarantees that $H_{\rm int}$ is real and well-defined. Replacing $k\rightarrow k_{i_1}$, $k_1 \rightarrow -k_{i_2}$, $k_2 \rightarrow k_{i_3}$, and $k_3 \rightarrow {k_{i_4}}$, and making use of $\sigma_{-k_{i_2}} = \sigma^*_{k_{i_2}}$, one can rewrite $H_{\rm int}$ in an equivalent ``symmetric" form as
\begin{equation}
H_{\rm int} = \frac{1}{4}\sum_{k_{i_1} + k_{i_2} = k_{i_3} + k_{i_4}} V_{-k_{i_1}, -k_{i_2}, k_{i_3}, k_{i_4}} \sigma^*_{k_{i_1}}\sigma^*_{k_{i_2}}\sigma_{k_{i_3}}\sigma_{k_{i_4}}.
\label{4W+h} 
\end{equation}
In the above, the quartic terms $\sigma^*_{k_{i_1}}\sigma^*_{k_{i_2}}\sigma_{k_{i_3}}\sigma_{k_{i_4}}$ represent the complex amplitudes of individual four-wave interaction events (scatterings, decays and mergers all count), the asterisk marks the falling waves, and we have omitted the Kronecker delta $\delta_{-k_{i_1}-k_{i_2}+k_{i_3}+k_{i_4},0}$ for simplicity. 

The fact that $H_{\rm int}$ collects both scatterings and decays via $\omega_{{k}_i} = \omega_i (|{k}_i|)$ has important physics implications. Indeed, it is shown theoretically \cite{Berman84a} and confirmed numerically \cite{Berman84b} that a certain amount of decays taking place is actually necessary in order for inelastic scatterings to occur. When the scatterings couple an increased number of waves (meaning nonlinearity exceeds a certain critical level), the system of interacting waves obeying Eqs.~(\ref{R4a})$-$(\ref{R4b}) naturally (without tuning of parameters) transits into a stochastic state, in which state it develops statistical, rather than deterministic, properties. It is this transition into a stochastic state that underlies the occurrence of ``turbulence" in large systems of interacting modes. In what follows, it is tacitly assumed that the necessary conditions \cite{Berman84b} for the stochastic instability to come into play have been satisfied, thus paving the way for the random dynamics \cite{Sagdeev,Zaslavsky}, and to a theoretical description in terms of the probability density function (Secs.~V and~VI).    

If one wants to single out the effect of four-wave interactions on the dynamics of field spreading (or if three-wave interactions are forbidden by the dispersion relation), then the procedure is to substitute~(\ref{4W}) into Eq.~(\ref{H0}) (in place of the three-wave $H_{\rm int}$) and apply the canonical equations~(\ref{Can}), from which the following equations of motion for the complex amplitudes ${\sigma_k}$ may be deduced:
\begin{equation}
\dot{\sigma}_k = i\omega_k\sigma_k + iV^*\sigma_{k_1} \sigma_{k_2} \sigma_{k_3},
\label{DEM4} 
\end{equation}
where we have denoted $V \equiv V_{k, -k_1, -k_2, -k_3}$ and $V^* \equiv V_{-k, k_1, k_2, k_3}$. Switching the indexes in Eq.~(\ref{DEM4}) and remembering that $\sigma_{-k} = \sigma_k^*$, one gets the dynamical equations for ${\sigma_{k_1}}$, i.e., 
\begin{equation}
\dot{\sigma}_{k_1} = i\omega_{k_1}\sigma_{k_1} - iV\sigma_{k} \sigma_{k_2}^* \sigma_{k_3}^*,
\label{DEM4-1} 
\end{equation}
and similarly for ${\sigma_{k_2}}$ and ${\sigma_{k_3}}$. The rate of field spreading is obtained as $R\simeq \rho |V^*|^2|\sigma_{n_1} \sigma_{n_2} \sigma_{n_3}|^2$, leading to [cf. Eq.~(\ref{Rate})]
\begin{equation}
R \simeq \rho |V|^2|\sigma_n|^6\simeq \rho |V|^2/(\Delta n)^3,
\label{Rate-ext} 
\end{equation}
where $\rho$ is a coefficient, $|\sigma_n|^2 \simeq 1/\Delta n$ in conformity with the conservation law in Eq.~(\ref{TP}), and we have assumed that $\Delta n$ is as large as to guarantee $R \ll \omega_k$ for the majority of $k$'s. Combining Eq.~(\ref{Rate-ext}) with Fermi's golden rule $R = d\Delta n/ dt$, one gets
\begin{equation}
d\Delta n / dt = \rho |V|^2/(\Delta n)^3, 
\label{Rate2-ext} 
\end{equation}
from which $(\Delta n)^4 = 4\rho |V|^2 t$. The latter equation corresponds to a subdiffusive spreading for $t\rightarrow+\infty$, i.e., 
\begin{equation}
(\Delta n)^2 = (4\rho)^{1/2} |V| t^{1/2}.
\label{SpL-ext} 
\end{equation}
The subdiffusive scaling law in Eq.~(\ref{SpL-ext}) is a familiar one. It is found in quantum chaotic dynamics, where it characterizes asymptotic spreading of a quantum wave packet in the nonlinear Schr\"odinger lattices with disorder (e.g., Refs. \cite{Many,Ivan,QW,PRE17,PRE19,PRE23}). This correspondence with quantum chaos is no surprise as we have based our model on Fermi's golden rule [Eq.~(\ref{Rate0})], with that justification that the interactions are resonant. If, instead of the golden rule, one applies the random-phase approximation as of Refs. \cite{PS,EPL,PRE14}, then a different scaling law is predicted for $t\rightarrow+\infty$, i.e., $(\Delta n)^2 \propto t^{2/5}$. We discount this scaling law here.   

\subsection{Exit-time distribution} 

The subdiffusive scaling laws in Eqs.~(\ref{SpL}) and~(\ref{SpL-ext}) correspond to a non-Markovian spreading process with exit-time statistics. The demonstration uses the idea of clustering of unstable modes in phase space (Refs. \cite{Iomin,PRE17}). Mathematically, it is convenient to unify the spreading laws in Eqs.~(\ref{Rate2}) and~(\ref{Rate2-ext}) by defining 
\begin{equation}
d\Delta n / dt = A /(\Delta n)^{2s+1}, 
\label{RR} 
\end{equation}
where the switcher $s$ takes the value $s=1$ for four-wave interactions and the value $s=1/2$ for three-wave interactions, and we have denoted $A = \rho |V|^2$ for simplicity. Integrating over time in Eq.~(\ref{RR}), one gets $(\Delta n)^{2s+2} = (2s+2)At$, from which
\begin{equation}
(\Delta n)^2 = [(2s + 2) A]^{1/(s+1)}t^{1/(s+1)}.
\label{Sub} 
\end{equation}
Differentiating both sides of Eq.~(\ref{RR}) with respect to time and eliminating the resulting $d\Delta n / dt$ on the right-hand side with the aid of the same Eq.~(\ref{RR}), one obtains 
\begin{equation}
\frac{d^2}{dt^2}\Delta n = -\frac{(2s + 1)A^2}{(\Delta n)^{4s + 3}}.
\label{Grad+} 
\end{equation}
Finally, by rewriting the power-law function on the right-hand side of Eq.~(\ref{Grad+}) such that it takes the form of a gradient against $\Delta n$, one gets
\begin{equation}
\frac{d^2}{dt^2}\Delta n = - \frac{d}{d \Delta n} \left[- \frac{A^2 / 2}{(\Delta n)^{4s + 2}}\right].
\label{Grad} 
\end{equation}
Equation~(\ref{Grad}) is equivalent to the Newton equation of motion of a point particle of unit mass in the potential field 
\begin{equation}
W (\Delta n) = - \frac{A^2 / 2}{(\Delta n)^{4s + 2}},
\label{Poten} 
\end{equation}
where $\Delta n$ has the sense of position coordinate and characterizes the actual span of the field distribution. 

If $s=1$, then the potential function in Eq.~(\ref{Poten}) becomes 
\begin{equation}
W (\Delta n) = - \frac{A^2 / 2}{(\Delta n)^{6}}.
\label{Poten+} 
\end{equation}
The latter potential is known from molecular physics, where it quantifies the attractive interactions between atoms inside molecules (as a constituent of the Lennard-Jones potential \cite{Lennard}). Given this insight and the fact that the potential function in Eq.~(\ref{Poten}) has attractive character for any $s > 0$, one might arguably propose that the newly excited modes form clusters, or ``molecules" in phase space \cite{Iomin,PRE17}, where they will be effectively trapped \cite{PRE19} due to their nonlinear coupling.  

Multiplying both sides of Eq.~(\ref{Grad}) by the velocity $d \Delta n /dt$ and integrating the ensuing differential equation with respect to time, after a simple algebra one obtains
\begin{equation}
\frac{1}{2}\left[\frac{d}{dt} \Delta n\right]^2 - \frac{A^2 / 2}{(\Delta n)^{4s + 2}} = \Delta E,
\label{Ener} 
\end{equation}
where the first term on the left-hand side has the sense of kinetic energy of a particle, and the second term is its potential energy. 

More so, it is shown using Eq.~(\ref{RR}) that the kinetic energy in Eq.~(\ref{Ener}) compensates for the potential energy {\it exactly}, that is the total energy in Eq.~(\ref{Ener}) is {zero}, {i.e.}, $\Delta E = 0$. Furthermore, both the negative potential energy $W (\Delta n) = - A^2 / 2(\Delta n)^{4s + 2}$ and the positive kinetic energy $\frac{1}{2}(d \Delta n /dt)^2 = A^2 / 2(\Delta n)^{2(2s + 1)}$ vanish while spreading. These both decay as the $(4s + 2)$th power of the number of states and the ratio between them does {\it not} depend on width of the field distribution. 

The total energy being equal to zero implies that a particle with the equation of motion~(\ref{Grad}) is sitting on the separatrix $\Delta E = 0$. That means that the motion process of this particle could be particularly sensitive to perturbations \cite{Zaslavsky,Chirikov_UFN,ChV}. Such perturbations may have different physics origins\textemdash from thermal noise to imprecision in the initial conditions \cite{Zaslavsky}\textemdash though the main cause is arguably the neglect of higher-order interaction terms in an idealized three- or four-wave picture of interactions \cite{Falkovich}. Here, we assume, following Refs. \cite{PRE24,PRE17}, that the role of random factors can be accounted for using the energy parameter $T$, and we interpret this as the ``temperature" of thermal bath enveloping the separatrix. 

Adding thermal fluctuations to the Lennard-Jones model in Eq.~(\ref{Poten}) leads directly to a non-Poissonian distribution of exit times with the divergent mean, as we now proceed to show. 

In fact, the probability for a given mode to quit the cluster after it has traveled $\Delta n$ sites on it is given by the Boltzmann factor 
\begin{equation}
p (\Delta n) = \exp [W (\Delta n) / T],
\label{BF} 
\end{equation}
where $W(\Delta n)$ is the negative potential energy stemming from the equation of motion~(\ref{Grad}). Substituting $W (\Delta n)$ from Eq.~(\ref{Poten}), one writes
\begin{equation}
p (\Delta n) = \exp [- A^2 / 2 T (\Delta n)^{4s + 2}].
\label{Escape} 
\end{equation}
Taylor expanding the exponential function for $\Delta n \gg 1$, one gets 
\begin{equation}
p (\Delta n) \simeq 1 - A^2 / 2 T (\Delta n)^{4s + 2}.
\label{Expand} 
\end{equation}
The probability to remain (survive) on the cluster after $\Delta n$ space steps is $p^{\prime} (\Delta n) = 1-p(\Delta n)$, yielding 
\begin{equation}
p^{\prime} (\Delta n) \simeq A^2 / 2 T (\Delta n)^{4s + 2}.
\label{EscapePr} 
\end{equation}
Eliminating $\Delta n$ with the aid of Eq.~(\ref{Sub}), one obtains the probability to survive on the cluster after $\Delta t$ time steps  
\begin{equation}
p^{\prime} (\Delta t) \propto (\Delta t)^{-(2s+1) / (s+1)},
\label{Survive} 
\end{equation}
which is equivalent to the exit-time distribution 
\begin{equation}
\chi_\alpha (\Delta t) \propto (\Delta t)^{-(1+\alpha)}
\label{WT} 
\end{equation}
with $\alpha = s/(s+1) < 1$. Specifically, one finds $\alpha = 1/3$ for three-wave interactions ($s=1/2$) and $\alpha = 1/2$ for four-wave interactions ($s=1$). One sees that the integral 
\begin{equation}
\int_{\sim 1}^{\tau} \Delta t \chi_\alpha (\Delta t) d\Delta t \sim \int_{\sim 1}^{\tau} (\Delta t)^{-\alpha} d\Delta t \sim \tau^{1-\alpha} \rightarrow+\infty
\label{Div} 
\end{equation}
diverges for $\tau\rightarrow+\infty$, implying that the mean exit time is infinite for all $\alpha < 1$. 

\section{Special case}

A special case of the above theory is a situation when two waves with oppositely directed wave vectors $k_j$ and $
-k_j$ couple together to form a bound state with zero momentum. The process is similar to the generation of the Cooper pairs in superconductors \cite{Cooper}. An interesting regime occurs when such bound states can participate in triad interactions with high $\bf k$ running waves via the zero-frequency resonance (see a schematic illustration in Fig.~4, top) 
\begin{equation}
\omega_{k_i} =  \omega_0 + \omega_{k_i - 0},
\label{R0} 
\end{equation} 
where $k_i$ and $k_i -0$ are wave vectors of the running wave just before and after an interaction event involving a bound state, and $\omega_0 \rightarrow 0$ is the vanishing frequency of this state. A situation of the kind is found in tokamaks, where the three-wave resonance in Eq.~(\ref{R0}) is held responsible for the generation of zonal flows \cite{Korean,Zonal,White,Zonal06}. By zonal flows one means azimuthally symmetric band-like shear flows, which are ubiquitous phenomena in planetary atmospheres, oceans and the laboratory \cite{Large,Korean,Zonal,McIntyre}. In the context of tokamak plasma, zonal flows are zero-frequency electrostatic potential fluctuations with finite radial wave number. These zonal flows are driven exclusively by nonlinear interactions, which transfer energy from electrostatic micro-turbulence into large-scale drift-like motion of plasma particles with electrostatic ${\bf E} \times {\bf B}$ drift. Usually, such nonlinear interactions are three-wave triad couplings between two high $\bf k$ drift waves and one low $\bf k$ zonal flow excitation \cite{Zonal}. The importance of zonal flows in magnetic confinement fusion is that these flows help to reduce the levels of turbulence and turbulent transport by absorbing the free energy from the high $\bf k$ turbulent background. 

\begin{figure}
\includegraphics[width=0.51\textwidth]{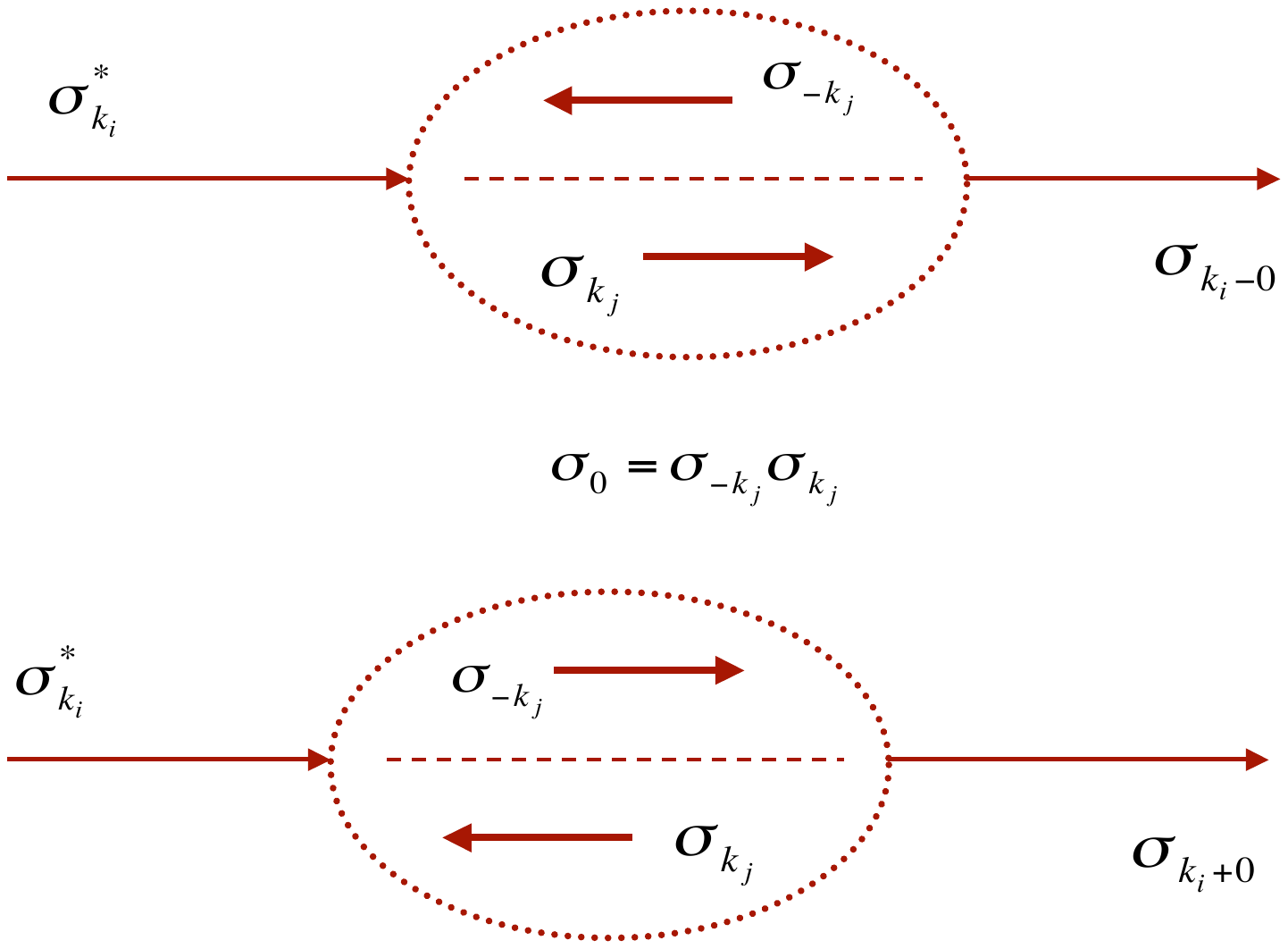}
\caption{\label{} The zero-frequency resonance. The oval structure (dotted line) represents a bound state between two counter-propagating waves with complex amplitudes $\sigma_{-k_j}$ and $\sigma_{k_j}$. Top: A running wave transferring energy to a bound state via the zero-frequency resonance. Bottom: The opposite process of a bound state transferring energy to a running wave.    
}
\end{figure}

The coupling process in Eq.~(\ref{R0}) corresponds to the interaction Hamiltonian 
\begin{equation}
H^\prime_{\rm int} = \sum_{k_i > 0} V_{-k_i, 0, k_i-0} \sigma^*_{k_i} \sigma_{0} \sigma_{k_i-0}\delta_{-k_i+0+k_i-0, 0},
\label{3W+a} 
\end{equation}
where $\sigma_0$ is the amplitude of a bound state with frequency $\omega_0\rightarrow 0$, the summation goes over all positive values of $k_i$ allowed by the dispersion relation, and we have kept the Kronecker delta for consistency with previous equations. 

The Hamiltonian in Eq.~(\ref{3W+a}) is joined by the Hamiltonian of exact opposite process  
\begin{equation}
H^{\prime\prime}_{\rm int} = \sum_{k_i > 0} V_{-k_i, -0, k_i+0} \sigma^*_{k_i} \sigma^*_{0} \sigma_{k_i+0}\delta_{-k_i-0+k_i+0, 0},
\label{3W+b} 
\end{equation}
which characterizes the decay of a bound state with frequency $\omega_0 \rightarrow 0$ via the zero-frequency resonance $\omega_{k_i} = - \omega_0 + \omega_{k_i + 0}$ (see Fig.~4, bottom). 

There is a subtlety here, however, and this refers specifically to fine structure of the bound states. Indeed, if the frequency of the bound states is about zero, i.e., $\omega_0 \rightarrow 0$, and if these states are formed by two counter-propagating waves having vanishing frequency each, then the resonance condition in Eq.~(\ref{R0}) can be satisfied \cite{PRE09} if only for a bound state as a whole as well as for each of the partial wave processes involved composing such a state. This is because the resonance in Eq.~(\ref{R0}) has finite width, which allows for a margin on admissible resonant frequencies. This finite spread over frequencies gives rise to an instability of the bound states\textemdash akin to stochastic instability of coupled nonlinear oscillators \cite{Zaslavsky,Chirikov_UFN}\textemdash which occurs along the separatrix $\omega _0 = 0$. The latter is a hypersphere in wave number space, with the center at zero wave number vector ${\bf k} = {0}$ and of radius $|{\bf k}_j|$ (see Sec. 7 of Ref. \cite{Stand}).  

More explicitly, if the interaction frequency $\omega_0 \rightarrow 0$, then the distance between resonances in vicinity of the separatrix behaves as $\delta\omega \sim \omega_0$, while the nonlinear resonance width \cite{Sagdeev} approaches zero in accordance with $\Delta \omega_{\rm NL} \propto \sqrt{\omega_0}$ \cite{PRE24} and for $\omega_0\rightarrow 0$ will be much larger than $\delta\omega$, i.e., $\Delta \omega_{\rm NL} \propto \sqrt{\omega_0} \gg \delta\omega \sim \omega_0$. The implication is that the zero-frequency resonance in Eq.~(\ref{R0}) is broad enough to cover the constituent wave processes over the entire bound state as $\omega_0 \rightarrow 0$. This zero-frequency instability, which is generic to isotropic systems with separatrix dynamics \cite{Sagdeev,ChV}, has been simulated numerically in Ref. \cite{Stand} in the context of surface waves in a fluid.  

Writing the amplitude of the bound states as $\sigma_{0} = \sigma_{-k_j}\sigma_{k_j}$, where $\sigma_{-k_j}$ and $\sigma_{k_j}$ are complex amplitudes of the constituent wave processes, and substituting into Eq.~(\ref{3W+a}), one may represent the interaction Hamiltonian $H^\prime_{\rm int}$ in an equivalent four-wave form
\begin{equation}
H^{\prime}_{\rm int} = \sum_{k_i > 0, k_j > 0} V_{-k_i, -k_j, k_j, k_i} \sigma^*_{k_i} \sigma^*_{k_j} \sigma_{k_j} \sigma_{k_i}\delta_{-k_i-k_j+k_j+k_i, 0},
\label{3W++} 
\end{equation}
where use has been made of $\sigma_{-k_j} = \sigma^*_{k_j}$. The Hamiltonian~(\ref{3W++}) is a partial case of the ``symmetric" Hamiltonian~(\ref{4W+h}), with that particularity that the summation in Eq.~(\ref{3W++}) goes over positive values of $k_i$ and $k_j$. 

One sees that the zero-frequency resonance is special in that it can be associated with either a three-wave process of the type given by Eqs.~(\ref{3W+a}) and~(\ref{3W+b}) or a four-wave process of the type given by Eq.~(\ref{3W++})\textemdash depending on whether the zero-frequency state $\sigma_{0} = \sigma_{-k_j}\sigma_{k_j}$ is accounted for as one single wave with zero momentum or two coupled waves with oppositely directed momenta. This duality has important statistical implications with regard to the dynamics of field spreading.  

In fact, if one wants to assess the asymptotic ($t\rightarrow+\infty$) dispersion of the wave field, then one needs to consider that the actual spreading rate $R = d\Delta n / dt$ is limited to the generation of new bound states by way of the zero-frequency resonance in Eq.~(\ref{R0}). In that regard, the zero-frequency state $\sigma_0$ behaves as one single wave, suggesting the interaction Hamiltonian in Eq.~(\ref{3W+a}) applies. The spreading law is therefore obtained from Eq.~(\ref{Sub}), where one demands $s=1/2$, leading to
\begin{equation}
(\Delta n)^2 \propto t^{2/3}.
\label{SpL+} 
\end{equation}   
On the other hand, focusing on exit-time statistics, one turns back on three-wave interactions and starts looking into four-wave processes instead, with that justification that it is this type of interaction process that limits the lifetime of the bound states (and therefore is most relevant for the exit-time distribution). It is at this point where the four-wave interaction Hamiltonian in Eq.~(\ref{3W++}) comes into play. In this approximation, the fact that $\sigma_0$ has internal structure is key, meaning the zero-frequency wave $\sigma_{0} = \sigma_{-k_j}\sigma_{k_j}$ is worth two interacting waves. It is understood that four-wave interactions result in a steeper distribution of exit times [see Eq.~(\ref{WT})] and as such will have an upper hand in determining the decay instability of the bound states. With this implication in mind, the distribution of exit times is inferred from Eq.~(\ref{WT}) by letting $s=1$ in $\alpha = s/(s+1)$, yielding
\begin{equation}
\chi_\alpha (\Delta t) \propto (\Delta t)^{-3/2}.
\label{WT+} 
\end{equation}  

Summarizing the above reasoning, it is noted that the presence of zero-frequency states results in a {\it mixed} statistics, when the asymptotic spreading law is three-wave-like, while the distribution of exit times is four-wave-like. 

In the context of tokamak plasmas, the spreading law in Eq.~(\ref{SpL+}) can describe the radial expansion of drift-wave turbulence by way of coupling to zonal flows. The subdiffusive character of Eq.~(\ref{SpL+}) suggests zonal flows may effectively suppress the radial transport. Indeed the fusion experiments demonstrate and direct computer simulations confirm that zonal flows can limit considerably the losses of hot thermonuclear plasma into the edge region (e.g., Refs. \cite{Lin,Kwon15,White,Zonal06,Militello}; Refs. \cite{Zonal,Terry} for reviews). This positive view should, however, be balanced by the fact that zonal flows simultaneously regenerate the turbulence through the decay process in Eq.~(\ref{3W+b}).      

An exciting result in the study of zonal flows in recent years has arguably been the discovery of quasiregular patterns of ${\bf E} \times {\bf B}$ flows dubbed the ${\bf E} \times {\bf B}$ staircase (or plasma staircase) \cite{DF2010,DF2015,DF2017,Horn2017}. Staircases are ubiquitous meso-scale dynamical structures characterized by narrow regions of localized gradient sharpening and of strong and lasting jets interspersed with broader regions of turbulent (typically, avalanching) transport (see Fig.~1 of Ref. \cite{PRE21}). Experimentally, ${\bf E} \times {\bf B}$ staircases are identified in a large variety of plasma parameters in ion drift-wave turbulence using correlation analysis of high-resolution fast-sweeping reflectometry (e.g., Ref. \cite{DF2015}). 

The plasma staircase exemplifies how a systematic organization of turbulent fluctuations may lead to the onset of strongly correlated flows on magnetic flux surfaces. Theoretically, the plasma staircase represents a synergetic cooperation between the transport by avalanches at the meso-scales and the spontaneously occurring zonal flows \cite{Nature,Korean}. The latter constitute a permeable localized barrier to avalanche propagation \cite{Beyer,Strugarek}. 

Focusing on the interaction Hamiltonian in Eq.~(\ref{3W++}), one may associate the zero-frequency states $\sigma_{0} = \sigma_{-k_j}\sigma_{k_j}$ with the different staircase jet zonal flows, where each such flow is labeled by a proper value of $k_j$. As each $k_j$ is excited on its own resonant magnetic flux surface, the interaction Hamiltonian~(\ref{3W++}) with the sum over $k_j$ predicts a qusiregular (grid-like) pattern, as observed \cite{DF2015,DF2017,Horn2017}. Developing these viewpoints, one might arguably propose that zonal flows naturally (under the overlap condition $\Delta \omega_{\rm NL} \gg \omega_0$) organize themselves into quasiregular spatio-temporal patterns that are seen in the experiment as the ${\bf E} \times {\bf B}$ staircase. In that regard, the subdiffusive scaling law in Eq.~(\ref{SpL+}) represents the radial dispersion of electrostatic drift waves that would result if the staircase structure contained an infinite number of jets.

\section{The comb model}

Perhaps the most important observation from the above analysis is the absence of a characteristic temporal scale of the spreading process. The unstable modes are trapped within disconnected clusters of states by the action of the Lennard-Jones potential in Eq.~(\ref{Poten}), and whether they do or do not quit the clusters is statistical, with a broad distribution of exit times. In this paradigm, a meaningful spreading occurs when the new modes are excited {outside} the existing clusters. Self-similarity implies that new excitations happen on a probabilistic basis and that the statistical distribution of waiting times pertaining to these excitations is actually the same as the exit-time distribution in Eq.~(\ref{WT}). 

In this Section, we analyze the expansion of the nonlinear field as a transport problem in comb geometry. The main idea here is that one can map the subdiffusive scaling laws in Eqs.~(\ref{SpL}) and~(\ref{SpL-ext}) onto a Dirac comb, and by doing so unveil the microscopic organization of the asymptotic spreading process. Technically, it is convenient to generalize the Dirac comb first, and to introduce a {\it comb space} by building the side branches {\it everywhere densely} along the backbone\textemdash instead of rising them at a fixed space step $\Lambda$ (see a schematic illustration in Fig.~5). For instance, the side branches could be placed at all rational numbers along the $x$ axis (because the rational numbers are everywhere dense in the set of real numbers). Because, on the other hand, the rational numbers can be made in one-to-one correspondence with the set of natural numbers \cite{Ewald}, a comb space with dense side branches is equivalent\textemdash as a mathematical set of points \cite{Stoll}\textemdash to the discrete comb in Eq.~(\ref{Pulse}). That means that a comb space with everywhere dense in the backbone side branches has the same cardinality \cite{Stoll} (contains the same infinite number of dynamical traps) as the analog discrete comb. It is understood that the asymptotic ($t\rightarrow+\infty$) dispersion law is determined by statistical distribution of waiting times spent within each such trap, and that all these traps are just identical copies of each other, suggesting one might rely on either model to assess the asymptotic dispersion. That said, a comb structure with dense side branches can be preferable to a discrete one as it simplifies the analysis behind the derivation of continuum equations (Sec.~V). From a fusion perspective, the notion of comb space with dense side branches is directly relevant to characterize the common properties of propagation of drift wave type waves in a tokamak as such waves get attracted by rational magnetic flux surfaces \cite{Garbet,Horton}. At the same time, the equivalence between the two formulations (discrete {versus} dense) facilitates the association of combs with zonal flows and staircases (Sec.~VI\,D). The comb model is formulated as follows.    

\begin{figure}
\includegraphics[width=0.55\textwidth]{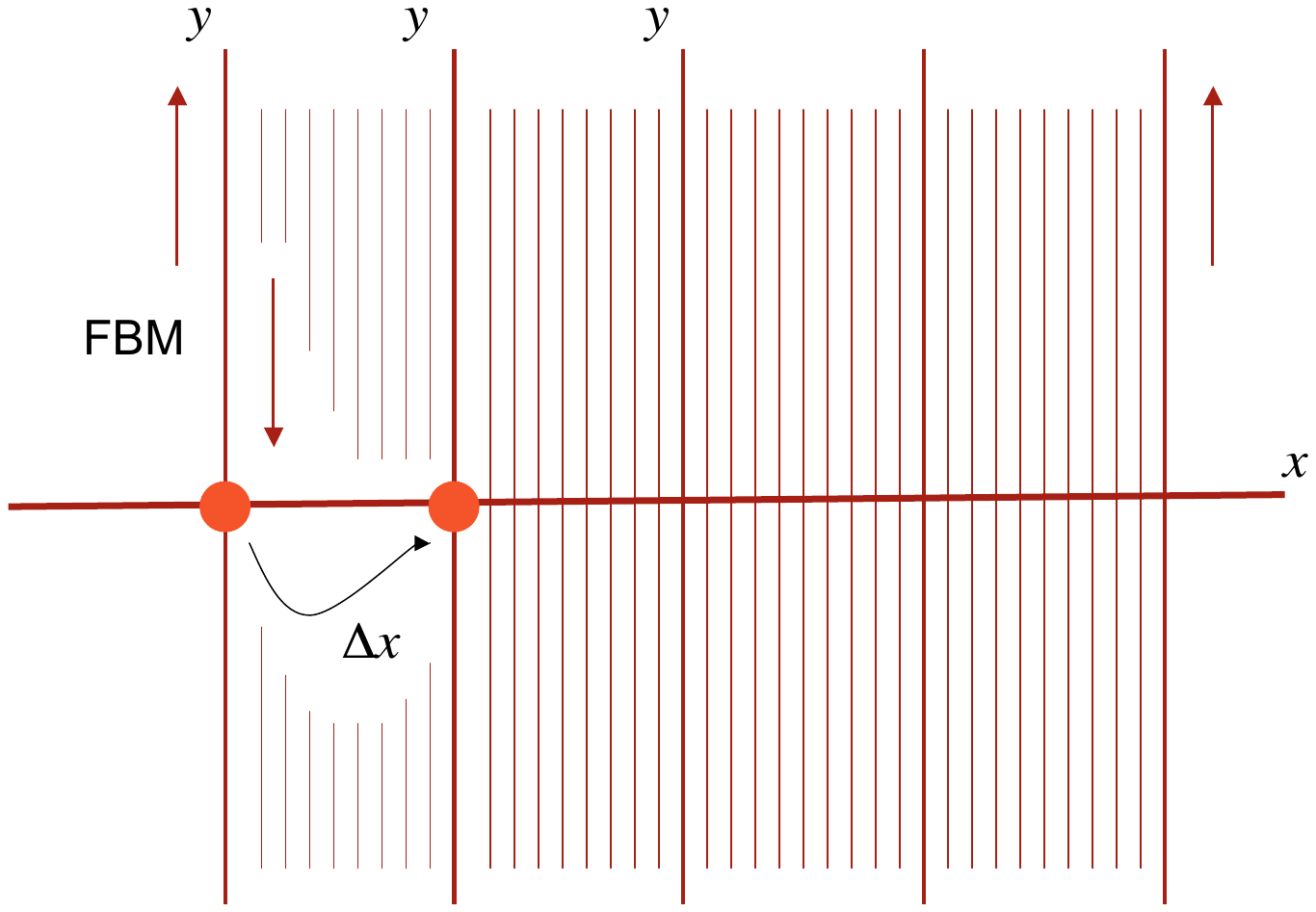}
\caption{\label{} The comb model. Side branches are shown as vertical lines and are assumed to be everywhere dense in the backbone (the horizontal line). The red circles represent the initial (left) and final (right) positions of the random walker performing a jump of the length $\Delta x$ along the backbone (marked by the coordinate $x$). Vertical arrows in the $y$ direction represent the fractional Brownian motion (FBM) in side branches.    
}
\end{figure}

Consider a comb space with everywhere dense in the backbone side branches as in Fig.~5. At time $t=0$, a motion process is initiated along an arbitrarily chosen side branch starting from an initial position at the backbone. This starting position, which is marked as ($x=0,\ y = 0$), is then used to set up a coordinate system in the entire comb space. In particular, we define $x$ and $y$ to be the position coordinates along the backbone and in side branches, respectively. The initial velocity is along $y$, if up- or downward, unimportantly.   

Further concerning the motion process, we take it to be fractional Brownian motion, or FBM. By FBM\textemdash dating back to Mandelbrot and van Ness \cite{Ness}\textemdash one means a Gaussian motion process such that the density of the probability to find a particle (random walker) at time $t$ at the distance $y$ away from the starting position is given by 
\begin{equation}
f (y, t) = (4\pi K_\beta t^\beta)^{-1/2} \exp [-y^2 / (4K_\beta t^\beta)], 
\label{FBM} 
\end{equation} 
where $0 < \beta \leq 2$ is the exponent of FBM. FBM is distinguished by the fact that it is the only self-similar Gaussian process with stationary increments (Ref. \cite{Ness}; Ref. \cite{Andrey} for a more recent discussion). The behavior of FBM is antipersistent for $0 < \beta < 1$ and persistent for $1 < \beta \leq 2$ \cite{Mandel}. If $\beta = 1$, then FBM reduces to normal diffusion, which is neither persistent nor antipersistent. Often the exponent of FBM is written as $\beta = 2H$, where $H$ is the Hurst exponent known from the time-series analyses \cite{Mandel,Feder}. Using $H$, one writes the mean-squared displacement along $y$ as  
\begin{equation}
\langle (\Delta y)^2(t)\rangle \propto t^{2H}, 
\label{SS} 
\end{equation} 
which is obtained straightforwardly as the second moment of the probability density function in Eq.~(\ref{FBM}).  

Focusing on tokamak applications, we associate the actual $H$ value with a competition between regular convection by the ${\bf E}\times {\bf B}$ drift and the trapping effect due to electrostatic micro-turbulence within the ${\bf E}\times {\bf B}$ flow. In this respect, the regular (convective) part favors long-time persistent behavior, with $1/2 < H \leq 1$, while the microscopic turbulence part imposes anti-persistent dynamics, characterized by $0 < H < 1/2$. The results, presented below, indicate $H \leq 1/2$ (i.e., the turbulent part dominates).   

Setting $y=0$ in Eq.~(\ref{FBM}), one obtains the probability to return to the starting position after $\Delta t$ time steps, i.e.,      
\begin{equation}
p_{y=0} (\Delta t) \propto (\Delta t)^{-H},
\label{STP} 
\end{equation} 
where the relation $H = \beta / 2$ has been used. In particular, if $H = 1/2$, then $p_{y=0} (\Delta t) \propto (\Delta t)^{-1/2}$. 

Next turning to the backbone transport, our assumptions are as follows. Any time the FBM walker returns to the backbone at $y=0$, it either jumps to another side branch some $\Delta x$ spatial steps away (with the probability $q$), or remains at the current side branch (with the probability $1-q$). If the walker does change the branch on which it is currently sitting, then it restarts from the same position $y=0$ on the new side branch, and the process repeats itself. These settings are illustrated schematically in Fig.~5. We shall assume for simplicity, without losing in generality, that the probabilities to jump and to remain are both equal to $q=1-q=1/2$, so the choice is random. Also we assume equal probabilities to jump along or against the $x$ axis, meaning there is no bias along the backbone. Finally, we define that the probability density to choose a new side branch as far away as $\Delta x$ spatial steps decreases with the number of steps as 
\begin{equation}
p_{\mu} (\Delta x) \sim A_\mu |\Delta x|^{-(1+\mu)}.
\label{PLF} 
\end{equation} 
That is, $p_{\mu} (\Delta x)$ is a decaying power-law function of $|\Delta x|$. In the above, $\mu$ is the exponent of the power law, $A_\mu$ is a normalization parameter, and we have tacitly assumed that the jump lengths may span over a broad range of spatial scales. The implication is that the jumps need not occur between the neighboring side branches only (in the discrete formulation), but there is a wide choice instead, though with a decaying probability density to jump onto a more distant side branch. Concerning the $\mu$ value, we restrict ourselves to the interval $1 < \mu < 2$. With this setting, the power-law distribution in Eq.~(\ref{PLF}) is L\'evy-stable for $|\Delta x| \gg 1$ \cite{Gnedenko,Bouchaud}. The case $0 < \mu < 1$, though mathematically similar, is not considered here. Yet so, we include the limiting case $\mu = 2$, which is understood as the normal (Gaussian) distribution of the jump lengths (because the normal distribution is the upper bound on L\'evy-stable distributions for $\mu\rightarrow 2$). Denoting the variance of the normal distribution as $\lambda^2$, we define 
\begin{equation}
\lim_{\mu\rightarrow 2} p_{\mu} (\Delta x) = \exp(-|\Delta x|^2/2\lambda^2).
\label{Norm} 
\end{equation}
Remark that the normal distribution in Eq.~(\ref{Norm}) imposes a characteristic jump length $\Delta x \sim \lambda$ in the $x$ direction. This contrasts the L\'evy statistical case, with $1 < \mu < 2$, in which case the distribution of the jump lengths is scale free. As is well-known \cite{Klafter,Chechkin,Bouchaud}, the L\'evy-stable distribution with $1 < \mu < 2$ generates L\'evy flights.     

The comb model described above is a generalization to L\'evy statistics of the random-walk model considered previously by Weiss and Havlin \cite{WeHa86}. In their model, the motion process in side branches is taken to be normal diffusion, not FBM. Also Weiss and Havlin assume a characteristic jump length along the backbone, which is set to be the period of a Dirac comb. In our notation, the model of Ref. \cite{WeHa86} is reproduced for $H = 1/2$, $\beta = 1$, and $\mu = 2$, leading to a waiting-time distribution $\chi_\alpha (\Delta t) \propto (\Delta t)^{-3/2}$. The analysis, presented below, suggests the model of Weiss and Havlin can characterize the spreading dynamics driven by four-wave interactions. Yet so, it appears that a simple comb model with the characteristic jump length fails to include if only triad interactions in Eq.~(\ref{3W}) as well as the zero-frequency resonance in Eq.~(\ref{R0}), in which cases one needs to introduce a comb space with everywhere dense in the backbone side branches, and a L\'evy-stable distribution of the jump lengths for $|\Delta x| \gg 1$.

\section{Transport equations}

A transport model for the random walk in comb geometry is obtained by joining together the Gaussian diffusion equation \cite{Ness,Mandel} for FBM in side branches and the space-fractional kinetic equation \cite{Klafter,Rest,Chechkin} for L\'evy flights along the backbone. Because, by the assumptions made, a L\'evy flight in the $x$ direction may only occur when FBM crosses the backbone at $y = 0$, we may write
\begin{equation}
\frac{\partial}{\partial t} f(x,y,t) = \left[K_\beta t^{\beta - 1}\frac{\partial^2}{\partial y^2} + \delta (y)K_\mu\frac{\partial^\mu}{\partial |x|^\mu}\right] f(x,y,t).
\label{KEK} 
\end{equation}
Here, $f = f(x,y,t)$ is the probability density to find the random walker at time $t$ at point ($x, y$) in the comb space, $\delta (y)$ is the Dirac delta-function, $K_\beta$ and $K_\mu$ are coefficients of the transport process,
\begin{equation}
\frac{\partial^\mu}{\partial |x|^\mu} f (x,y,t) = \frac{1}{\Gamma_\mu}\frac{\partial^2}{\partial x^2} \int_{-\infty}^{+\infty}\frac{f (x^\prime,y,t)}{|x-x^\prime|^{\mu - 1}} dx^\prime
\label{Def+} 
\end{equation} 
is the Riesz fractional derivative \cite{Podlubny,Samko} of order $\mu$, $1 < \mu < 2$ is the L\'evy index, and $\Gamma_\mu = -2\cos(\pi\mu/2)\Gamma(2-\mu)$ is a normalization parameter. The latter parameter ensures smooth crossover to $\partial^2/\partial x^2$ of the fractional operator $\partial^\mu/\partial |x|^\mu$ as $\mu\rightarrow 2$, i.e., $\lim_{\mu\rightarrow 2} {\partial^{\mu}}/\partial {|x|^{\mu}} = {\partial^{2}}/\partial {x^{2}}$. Note that $\Gamma_\mu\rightarrow+\infty$ for $\mu\rightarrow 2$ from below. Note, also, that the Riesz fractional derivative $\partial^\mu/\partial |x|^\mu$ is an integro-differential operator for all $0 < \mu < 2$ (in contrast to a conventional derivative of integer order). As such, it incorporates some nonlocality of spreading dynamics (because the probability densities at points $x$ and $x^\prime$ appear to be long-range correlated through the slowly decaying kernel $\propto 1/|x - x^\prime|^{\mu - 1}$). For $\mu \rightarrow 2$, the nonlocal features vanish by way of ${\partial^{\mu}}/\partial {|x|^{\mu}} \rightarrow {\partial^{2}}/\partial {x^{2}}$, giving rise to a local (in the sense of the central limit theorem \cite{Gnedenko}) asymptotic transport process. If $\mu\rightarrow 1$, then the Riesz fractional derivative~(\ref{Def+}) reduces to \cite{Mainardi}
\begin{equation}
\frac{\partial^\mu}{\partial |x|^\mu} f (x,y,t) \rightarrow - \frac{1}{\pi} \frac{\partial}{\partial x} \int_{-\infty}^{+\infty}\frac{f (x^\prime, y, t) }{x-x^\prime} dx^\prime,
\label{Hilbert} 
\end{equation} 
where the spatial derivative $\partial / \partial x$ acts on the Hilbert transform operator. The latter is defined by \cite{Rest}
\begin{equation}
\hat{\bf H} [f (x,y,t)] = \frac{1}{\pi}\int_{-\infty}^{+\infty}\frac{f (x^\prime, y, t) }{x-x^\prime} dx^\prime.
\label{Hilbert+} 
\end{equation} 
For $\mu\rightarrow 2$, the L\'evy-fractional Eq.~(\ref{KEK}) becomes
\begin{equation}
\frac{\partial}{\partial t} f(x,y,t) = \left[K_\beta t^{\beta - 1}\frac{\partial^2}{\partial y^2} + \delta (y)K_2\frac{\partial^2}{\partial x^2}\right] f(x,y,t),
\label{KEK+} 
\end{equation}
where $K_2$ characterizes diffusivity along the backbone. Letting $\beta = 1$, one obtains the Fokker-Planck equation in the comb space 
\begin{equation}
\frac{\partial}{\partial t} f(x,y,t) = \left[K_1 \frac{\partial^2}{\partial y^2} + \delta (y)K_2\frac{\partial^2}{\partial x^2}\right] f(x,y,t),
\label{KEK+N} 
\end{equation}
which does not contain the scaling factor $\sim  t^{\beta - 1}$ in front of the ${\partial^2}/{\partial y^2}$ term. 

We note in passing that the transport model in Eq.~(\ref{KEK+N}) with normal diffusion in side branches was considered by Arkhincheev and Baskin \cite{ArBa91} as a model of subdiffusive transport along Weiss and Havlin's comb \cite{WeHa86}. 

Note, also, that we write continuum equations for the probability density function $f = f (x, y, t)$, where $x$ is a real number. This is licit as we rise the side branches everywhere densely along the backbone, meaning any real position $x$ on the backbone can be approximated, as accurately as one likes, by a proper rational value \cite{Stoll}. 
     
A common feature among Eqs.~(\ref{KEK}),~(\ref{KEK+}) and~(\ref{KEK+N}) is the presence of singularity in the backbone term $\delta (y)K_\mu {\partial^\mu} f (x, y, t)/{\partial |x|^\mu}$. This singularity is represented by the Dirac delta-function, $\delta (y)$, and accounts for coupling between the transport processes along the backbone and in side branches. If $y\ne 0$, i.e., the random walker is outside the backbone, then the backbone term is cancelled out to zero, leaving back the familiar diffusion equation for FBM in the $y$ direction, i.e.,    
\begin{equation}
\frac{\partial}{\partial t} f(y,t) = K_\beta t^{\beta - 1}\frac{\partial^2}{\partial y^2} f(y,t).
\label{FBM+} 
\end{equation}
Based on this equation, one recovers the probability density in Eq.~(\ref{FBM}), from which the asymptotic dispersion law \cite{Mandel,Feder} for FBM can be inferred, i.e.,  
\begin{equation}
\langle (\Delta y)^2 (t) \rangle = \int_{-\infty}^{+\infty} (\Delta y)^2 f (\Delta y, t) d\Delta y \propto t^\beta,
\label{DIS} 
\end{equation}
where $\beta = 2H$ is the exponent of FBM, and $t\rightarrow+\infty$. 

If $y=0$, then instead of cancellation to zero one encounters a divergent behavior in the backbone term due to the Dirac delta-pulse. Technically, this is a problem, since the pulse function is non-analytical. To circumvent this difficulty, one may consider that the probability~(\ref{STP}) to return to the starting position is equivalent to a waiting-time distribution between consecutive steps of the random walk along the backbone, i.e.,       
\begin{equation}
\chi_H (\Delta t) \sim d p_{y=0} (\Delta t) / d\Delta t \propto (\Delta t)^{-(H+1)}.
\label{STP+} 
\end{equation}    
On account of this last distribution one may write, instead of Eq.~(\ref{KEK}), the effective 1D equation 
\begin{equation}
\frac{\partial}{\partial t} f(x,t) = {_0}{{D}}_t^{1-\beta/2} K_{\beta, \mu}\frac{\partial^\mu}{\partial |x|^\mu} f(x,t),
\label{KEK-RL} 
\end{equation}
where 
\begin{equation}
{_0}{{D}}_t^{1-\beta/2} f (x, t) = \frac{1}{\Gamma (\beta/2)}\frac{\partial}{\partial t}\int _{0}^{t} \frac{f (x, t^{\prime})}{(t - t^{\prime})^{1-\beta/2}}dt^\prime \label{R-L} 
\end{equation}
is the Riemann-Liouville fractional derivative \cite{Podlubny,Samko} of the order $1-\beta/2 = 1-H$. Indeed, it is shown \cite{Klafter,Sokolov} in a basic theory of CTRWs that the non-Poissonian distribution of waiting times in Eq.~(\ref{STP+}) leads directly to a time-fractional equation~(\ref{KEK-RL}) with the Riemann-Liouville derivative~(\ref{R-L}) on the right-hand side. This fractional derivative of the order $1-H$ is a consequence of the fact that the mean waiting time resulting from~(\ref{STP+}) is infinite, i.e., 
\begin{equation}
\int_{\sim 1}^{\tau} \Delta t \chi_H (\Delta t) d\Delta t \propto \tau^{1-H} \rightarrow+\infty,
 \label{Div2} 
\end{equation}
where $\tau\rightarrow+\infty$, and $0 < H < 1$.

We should stress that the model~(\ref{KEK-RL}) is {\it not} an equivalent of the original 2D model in Eq.~(\ref{KEK}), but an effective 1D reduction of this \cite{IomBas}, making it possible to avoid dealing with the singularity in the backbone term. This is achieved on the expense of introducing a special form of time differentiation using the Riemann-Liouville fractional operator~(\ref{R-L}). Mathematically, the Riemann-Liouville derivative~(\ref{R-L}) is analogous to the Riesz derivative in Eq.~(\ref{Def+}), though it uses a proper range of integration, with the lower limit set to $t=0$. This lower limit is dictated by a condition that the walk process starts at the time $t=0$. This way, the Riemann-Liouville operator incorporates the initial-value problem into the transport model. Remark that the transport equation~(\ref{KEK-RL}) can be rewritten in an equivalent form using the Caputo fractional derivative \cite{Klafter,Podlubny}\textemdash in that case the fractional differentiation over time goes to the left-hand side of the corresponding transport equation (where it replaces the $\partial/\partial t$ derivative), and has the order of fractional differentiation equal to $H = \beta / 2$. Because of this integro-differential character of fractional differentiation, the ensuing transport model proves to be non-Markovian, i.e., the current state of the transport process depends on the past states, with a long-time memory kernel.    

Using kinetic Eq.~(\ref{KEK-RL}), one obtains the fractional moments 
\begin{equation}
\langle |\Delta x|^\gamma (t) \rangle \propto t^{\gamma\beta/2\mu}
\label{Pseudo} 
\end{equation}  
of the $f = f (x, t)$ distribution, from which the scaling of the pseudo mean-squared displacement $\langle (\Delta x)^2 (t) \rangle$ versus time can be deduced for $t\rightarrow+\infty$, i.e., 
\begin{equation}
\langle (\Delta x)^2 (t) \rangle \equiv [\langle |\Delta x|^\gamma (t) \rangle]^{2/\gamma} \propto t^{\beta/\mu},
\label{Pseudo2} 
\end{equation}  
where $0 < \gamma < \mu \leq 2$. In the above $\langle\dots\rangle$ denotes the ensemble average, i.e., $\langle[\dots]\rangle = \int_{-\infty}^{+\infty} [\dots]dx$, where the integration is performed in infinite limits along the backbone. Note that we obtain the pseudo mean-squared displacement in Eq.~(\ref{Pseudo2}) by rescaling the fractional moment $\langle |\Delta x|^\gamma (t) \rangle$ as the direct calculation of the second moment of $f (x, t)$ yields a divergent result owing to the nonlocal character of L\'evy flights \cite{Klafter,Chechkin}.

Combining Eqs.~(\ref{DIS}) and~(\ref{Pseudo2}), one obtains a relationship between the respective dispersions in the $x$ and $y$ directions, i.e.,
\begin{equation}
\langle (\Delta x)^2 (t) \rangle \propto [\langle (\Delta y)^2 (t)]^{1/\mu}.
\label{COM} 
\end{equation}
This last equation shows that the diffusion on combs is inherently anisotropic, with a faster component along the side branches, and a slower component along the backbone (for $1 < \mu \leq 2$). 

\section{Transport equations continued}

The fractional transport model in Eq.~(\ref{KEK-RL}) characterizes the expansion of wave turbulence by inelastic wave-wave interactions for $t\rightarrow+\infty$. The asymptotic dispersion law in physical space is obtained by identifying the backbone coordinate $x$ with the radial spread $\Delta x$. The latter is proportional to the number of states $\Delta n$ for the reasons explained in the end of Sec.~II\,A, i.e., $\Delta x \propto \Delta n$. We have, with the aid of Eq.~(\ref{Pseudo2}), 
\begin{equation}
\langle (\Delta n)^2 (t) \rangle \propto t^{\beta/\mu},
\label{NoS} 
\end{equation}
where we have kept angle brackets to emphasize that the scaling in Eq.~(\ref{NoS}) derives statistically by taking moments of the probability density function. 

In what follows, we consider separately the three- and four-wave interaction cases, as well as the special case of the zero-frequency resonance: 

\subsection{Three-wave interactions}

In a three-wave process, the turbulence expansion law is given by the subdiffusive transport scaling in Eq.~(\ref{SpL}). Matching~(\ref{SpL}) to~(\ref{NoS}) yields $\beta/\mu = 2/3$. The $\beta$ value is obtained by comparing the waiting-time distributions in Eqs.~(\ref{WT}) and~(\ref{STP+}), leading to $\alpha = H$, from which $\beta = 2\alpha$. Letting $s=1/2$ in $\alpha = s/(s+1)$, one gets $\alpha = 1/3$, $H=1/3$, and $\beta = 2/3$. Returning to Eq.~(\ref{NoS}), one infers $\mu = 1$. The latter exponent renders the limiting form~(\ref{Hilbert}) to the Riesz fractional operator in Eq.~(\ref{Def+}). Equation~(\ref{Hilbert}) tells us that the transport along the backbone is dominated by Cauchy-L\'evy flights \cite{Rest,Chechkin}, with the jump length distribution  
\begin{equation}
\chi_{\mu} (\Delta x) \sim A_1 (\Delta x)^{-2}
\label{PLF+} 
\end{equation} 
consistently with the probability density in Eq.~(\ref{PLF}). The asymptotic transport equation in the $x$ direction is deduced from the general Eq.~(\ref{KEK-RL}), where one employs $\beta = 2/3$ and $\mu \rightarrow 1$. Using~(\ref{Hilbert}), one is led to a nonlocal (in the sense of the generalized central limit theorem \cite{Gnedenko,Bouchaud}) transport model
\begin{equation}
\frac{\partial}{\partial t} f (x, t) = -\frac{1}{\pi}\,_{0}{D}_t^{2/3} \left[K_{2/3, 1}\frac{\partial}{\partial x} \int_{-\infty}^{+\infty}\frac{f (x^\prime, t)}{x-x^\prime} dx^\prime\right].
\label{Hil} 
\end{equation}
The nonlocal character of Eq.~(\ref{Hil}) is clear from the convolution operator on the right-hand side, which integrates the slowly decaying kernel $\propto 1 / (x - x^\prime)$ in infinite limits. 

In fusion literature, there exists an impressive evidence that the spillover of turbulence into stable regions could be nonlocal, with a wealth of data encompassing edge turbulence \cite{JJR,Basu1,Naulin,Naulin07,Hariri}, energetic particles \cite{Zonca06,Heid,Zonca15,Zonca_RMF} and edge-SOL coupling \cite{Short,Korean,Nature,Singh}. Our results indicate that nonlocal behavior is produced naturally through inelastic triad interactions in a multi-wave Hamiltonian system with the interaction Hamiltonian~(\ref{3W})\textemdash consistently with the fractional transport Eq.~(\ref{Hil}) and the Hilbert transform operator~(\ref{Hilbert+}).    

\subsection{Four-wave interactions}

Mathematically, the four-wave interaction case is similar to triad interactions considered previously, yet it leads to a somewhat different kinetic description, as we now proceed to show. 

Combining the scaling laws in Eqs.~(\ref{SpL-ext}) and~(\ref{NoS}), one is led to $\beta/\mu = 1/2$. The distribution of waiting times is deduced from Eq.~(\ref{WT}), where the $\alpha$ value is obtained by letting $s=1$ in $\alpha = s/(s+1)$, from which $\alpha = 1/2$. Comparing to~(\ref{STP+}), one finds $H=1/2$ and $\beta= 2H = 1$. With the aid of Eq.~(\ref{Pseudo2}) one gets $1/\mu = 1/2$, i.e., $\mu = 2$. The transport model in Eq.~(\ref{KEK-RL}) becomes   
\begin{equation}
\frac{\partial}{\partial t} f(x,t) = {_0}{{D}}_t^{1/2} K_{1, 2}\frac{\partial^2}{\partial x^2} f(x,t),
\label{KEK-RL+} 
\end{equation}
which contains the fractional derivative over time, but not over the space variable. 

Note that the condition $\mu = 2$ imposes a characteristic jump length in the $x$ direction via a crossover to Gaussian statistics in the limit $\mu \rightarrow 2$ [see Eq.~(\ref{Norm})]. That means that the transport model in Eq.~(\ref{KEK-RL+}) satisfies the defining conditions of the central limit theorem \cite{Gnedenko,Kampen} and in this sense is local, in contrast to the nonlocal model in Eq.~(\ref{Hil}). This absence of nonlocal features is a distinct feature of four-wave interactions \cite{PRE24}. Yet, the model in Eq.~(\ref{KEK-RL+}) is non-Markovian (includes long-time correlations) owing to the Riemann-Liouville derivative ${_0}{{D}}_t^{1/2}$ on the right-hand side. 

The fundamental solution or Green's function of the time-fractional Eq.~(\ref{KEK-RL+}) can be expressed in terms of the Fox $H$-function (Appendix B of Ref. \cite{Klafter}) as 
\begin{equation}\label{Fox} 
f (x, t) = \frac{1}{\sqrt{4\pi K_{1, 2} t^{1/2}}}
H_{1,2}^{2,0}\left[\frac{x^2}{4 K_{1, 2} t^{1/2}}\left|\begin{array}{c c}
(3/4, 1/2)\\
 (0,1),(1/2,1)\end{array}\right.\right],
\end{equation}
from which the subdiffusive scaling $\langle (\Delta x)^2 (t) \rangle \propto t^{1/2}$ can be inferred. 

It is understood that the subdiffusive transport scaling $\langle (\Delta x)^2 (t) \rangle \propto t^{1/2}$ occurs as a result of delta-coupling between the degrees of freedom in the original 2D Fokker-Planck model in Eq.~(\ref{KEK+N}). The effect this coupling has on asymptotic dynamics is that there is a waiting-time distribution between consecutive steps of the random walk along the backbone, which is mathematically equivalent to the fractional transport model in Eq.~(\ref{KEK-RL+}). It is instructive to demonstrate how the original 2D model with coupling\textemdash which does not contain fractional derivatives\textemdash produces the same subdiffusive transport scaling $\langle (\Delta x)^2 (t) \rangle \propto t^{1/2}$ as the fractional model in Eq.~(\ref{KEK-RL+}). We have collected this demonstration in Appendix A.    
      
\subsection{Special case}

Turning to the zero-frequency resonance (i.e., the special case discussed in Sec.~III), one combines the dispersion of a three-way interaction process, i.e., $\langle(\Delta n)^2(t)\rangle \propto t^{2/3}$, with the distribution of waiting times dictated by four-wave interactions, i.e., $\chi_\alpha (\Delta t) \propto (\Delta t)^{-3/2}$. 

More explicitly, by matching the general dispersion law in Eq.~(\ref{NoS}) to the three-wave transport scaling in Eq.~(\ref{SpL+}) one gets $\beta/\mu = 2/3$. On the other hand, by comparing the waiting-time distributions in Eqs.~(\ref{WT+}) and~(\ref{STP+}) one obtains $H=1/2$, $\alpha = 1/2$ and $\beta = 1$, from which $1/\mu = 2/3$ and $\mu = 3/2$. The distribution of the jump lengths in Eq.~(\ref{PLF}) becomes
\begin{equation}
\chi_{\mu} (\Delta x) \sim A_{3/2} |\Delta x|^{-5/2},
\label{PLF+M} 
\end{equation} 
where $A_{3/2}$ is a normalization parameter. The resulting asymptotic transport equation is inferred from bi-fractional Eq.~(\ref{KEK-RL}) by letting $\beta = 1$ and $\mu = 3/2$, leading to 
\begin{equation}
\frac{\partial}{\partial t} f(x,t) = {_0}{{D}}_t^{1/2} K_{1, 3/2}\frac{\partial^{3/2}}{\partial |x|^{3/2}} f(x,t).
\label{KEK-SP} 
\end{equation}
This equation is different from Eq.~(\ref{KEK-RL+}) in that it contains the fractional derivatives with regard to {\it both} time $t$ and the space variable $x$ [similarly to Eq.~(\ref{Hil}) with Cauchy-L\'evy flights]. 

\subsection{The ${\bf E} \times {\bf B}$ staircase as a Dirac comb}

The mapping of the ${\bf E} \times {\bf B}$ staircase (Sec.~III) onto a Dirac comb is built in slab geometry as in Fig.~6. For symmetry reasons, the slab is drawn in proximity to the neutral line in the poloidal cross-section. The backbone coordinate $x$ is in radial direction. The jet zonal flows are represented by the teeth of the comb. The side coordinate $y$ mimics the poloidal coordinate. We adopt a level of idealization according to which $y$ is a Pythagorean coordinate, not a cyclic one. The cyclic case is mathematically similar, though is not considered here. Following Ref. \cite{PRE18}, the period scale $\Lambda$ between neighboring teeth is evaluated as the electrostatic Rhines length \cite{Naulin2}, i.e., $\Lambda \simeq \Lambda_{\rm Rh}$. Similarly to its celebrated fluid analog \cite{McIntyre}, the electrostatic Rhines length $\Lambda_{\rm Rh}$ designates the spatial scale separating vortex motion from drift wave-like motion. As such, it scales as the square root of the fluid (drift) velocity, i.e., $\Lambda_{\rm Rh} \simeq \sqrt{|{\bf E} \times {\bf B}|}$, where $\bf E$ is the radial electric field, and $\bf B$ is the toroidal magnetic field. Note that we employ a discrete comb as in Eq.~(\ref{Pulse}), which is more suitable to represent the ${\bf E} \times {\bf B}$ staircase, owing to a well-defined spacing between the jets, i.e., $\Lambda \simeq \Lambda_{\rm Rh}$. 

\begin{figure}
\includegraphics[width=0.51\textwidth]{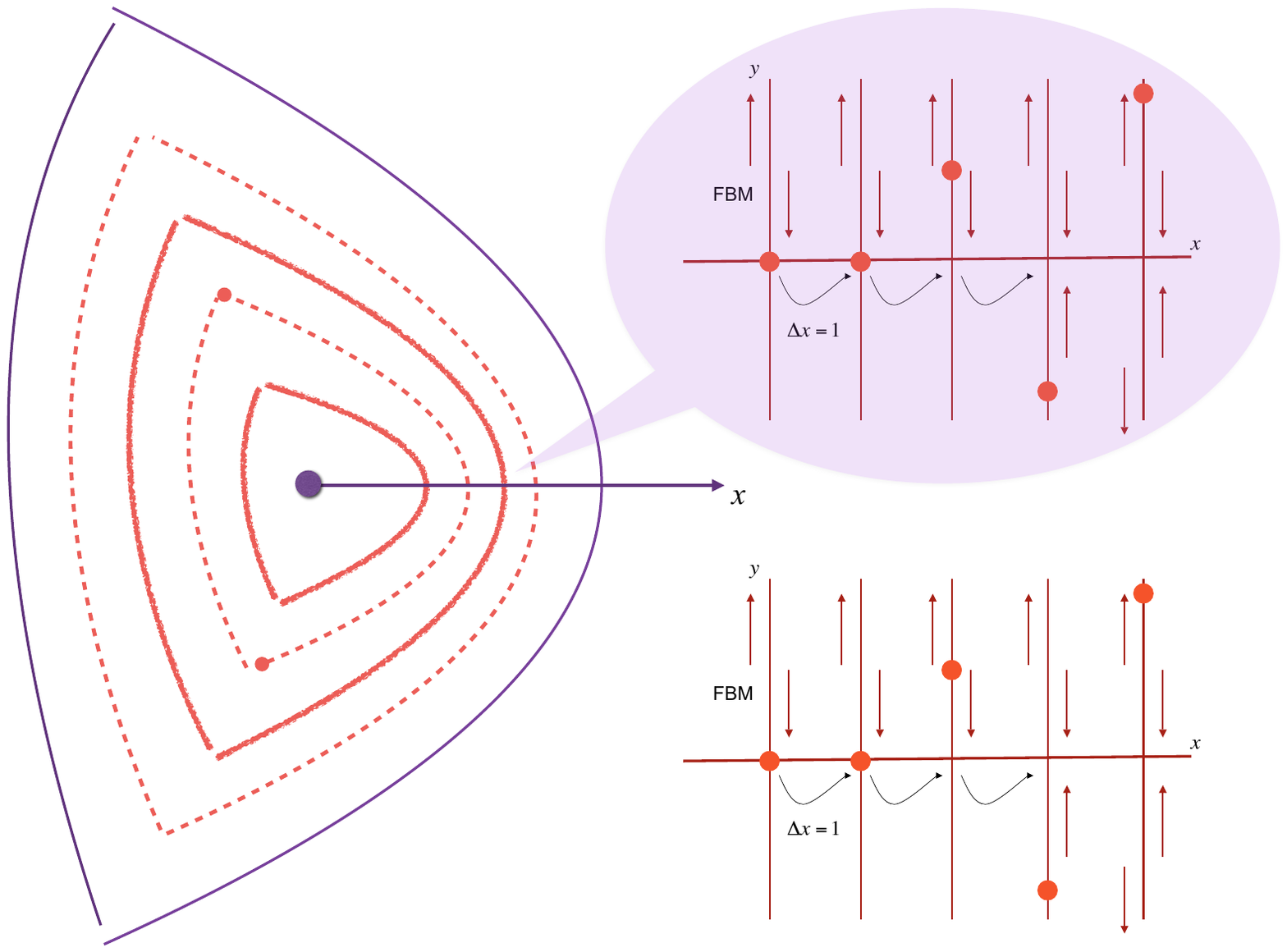}
\caption{\label{} The ${\bf E \times \bf B}$ staircase as a Dirac comb. Left: a poloidal cross-section of magnetic flux surfaces in a tokamak. The flux surfaces are shown as alternating solid and dashed lines (red color). The plasma staircase occurs at a crossroads between the outer core and inner edge plasma and is enlarged at the top-right of the Figure. The same structure is reproduced at the bottom-right. Following Ref. \cite{PRE18}, one associates the jet poloidal flows with side branches of the Dirac comb. The backbone direction, marked by the coordinate $x$, represents the radial direction in a tokamak. A tracer particle (random walker, red circle) can jump from the side branch on which it is currently sitting to either a neighboring side branch (a situation depicted above) or bypass the neighboring branch to land at a more distant side branch, such that there is a distribution of the jump lengths in accordance with the L\'evy distribution in Eq.~(\ref{PLF}). Vertical arrows (both directions) represent the fractional Brownian motion (FBM) in side branches.        
}
\end{figure}

A dynamical model is obtained by assuming that the random walker may occasionally jump (with the probability $1/2$, as it comes across the backbone) from the side branch on which it is currently sitting to another side branch: to either a neighboring one (a situation depicted in Fig.~6) or a more distant one, with a L\'evy distribution of the jump lengths as in Eq.~(\ref{PLF}). These model assumptions find support in the direct experimental observation of plasma avalanches \cite{DF2015,Beyer} and the fact that the staircase jets operate as semi-permeable \cite{Horn2017,DF2017,Strugarek} transport barriers to radial transport, i.e., whether or not an avalanche is absorbed by a given jet is probabilistic. 

More so, the waiting-time distribution between consecutive jumps (consecutive emission-reabsorption events) is borrowed from Eq.~(\ref{WT+}) on account of the zero-frequency resonance~(\ref{R0}), i.e., $\chi_\alpha (\Delta t) \propto (\Delta t)^{-3/2}$. This last distribution, together with the distribution of the jump lengths in Eq.~(\ref{PLF+M}), leads directly to bi-fractional transport equation~(\ref{KEK-SP}), where the fractional derivatives over time $t$ and the spatial coordinate $x$ incorporate the signatures of non-Markovianity and nonlocality, respectively.   

One sees that the staircase dynamics is both long-time correlated ($\alpha = 1/2$) and nonlocal ($\mu = 3/2$). 
      
It is understood that the L\'evy distribution of the jump lengths in Eq.~(\ref{PLF}) implies a power-law distribution of plasma avalanches over their sizes \cite{PRE18,PRE21}, i.e., 
\begin{equation}
\chi_{\mu} (\ell) \sim A_\mu \ell^{-(1+\mu)},
\label{PLF-L} 
\end{equation} 
where by ``size" $\ell$ one means the radial distance traveled by an avalanche since its emission at one radial location and up to an eventual absorption at another location some $\ell$ radial steps away. 

In fusion studies, the size distribution of plasma avalanches across the ${\bf E} \times {\bf B}$ staircase has been measured numerically in computer simulations \cite{DF2017,Horn2017} of the Tore Supra plasma \cite{DF2015} using the \textsc{Gysela} code \cite{Sarazin}. The results from those simulations can be summarized by a Fr\'echet distribution, which is a special case of Weibull (or generalized extreme value) distribution with lower bound \cite{Haan}. For large (covering at least one staircase period or more) avalanches, having size $\ell \gtrsim \Lambda_{\rm Rh}$, a behavior compatible with the Pareto-L\'evy distribution has been observed \cite{DF2017,PRE21}, i.e.,
\begin{equation}
\chi_{\kappa} (\ell) \sim A_\kappa \ell^{-(1+1/\kappa)},
\label{PLF-W} 
\end{equation} 
where $\kappa$ is the fitting shape parameter, with the best fit found at $\kappa \simeq 0.67$ (in the notation of Ref. \cite{PRE21}, $\ell \equiv \Delta n$ and $\chi_\kappa \equiv \mathcal{F}_k$). Comparing to Eq.~(\ref{PLF-L}), one identifies the exponent $1/\kappa$ with the L\'evy index $\mu$, i.e., $\kappa = 1/\mu$. Using $\mu = 3/2$, one infers $\kappa = 2/3$, which almost precisely reproduces the numerical result $\kappa \simeq 0.67$ \cite{PRE21}. 

Here, for the reader's convenience we incorporate a plot from Ref. \cite{PRE21}, which reports the computed probability distribution $\mathcal{F}_k (\Delta n)$ versus the Tore Supra data and the approximations made. The plot, which is shown in Fig.~7, evidences on its right the optimal fitting shape parameter $\kappa$, with the extremum of fitting quality at $\kappa \simeq 0.67$. We envisage this close agreement with numerical simulations \cite{DF2017,Horn2017,PRE21} as an important milestone towards the validation of the comb model and the associated bi-fractional transport Eq.~(\ref{KEK-SP}).

\begin{figure}
\includegraphics[width=0.49\textwidth]{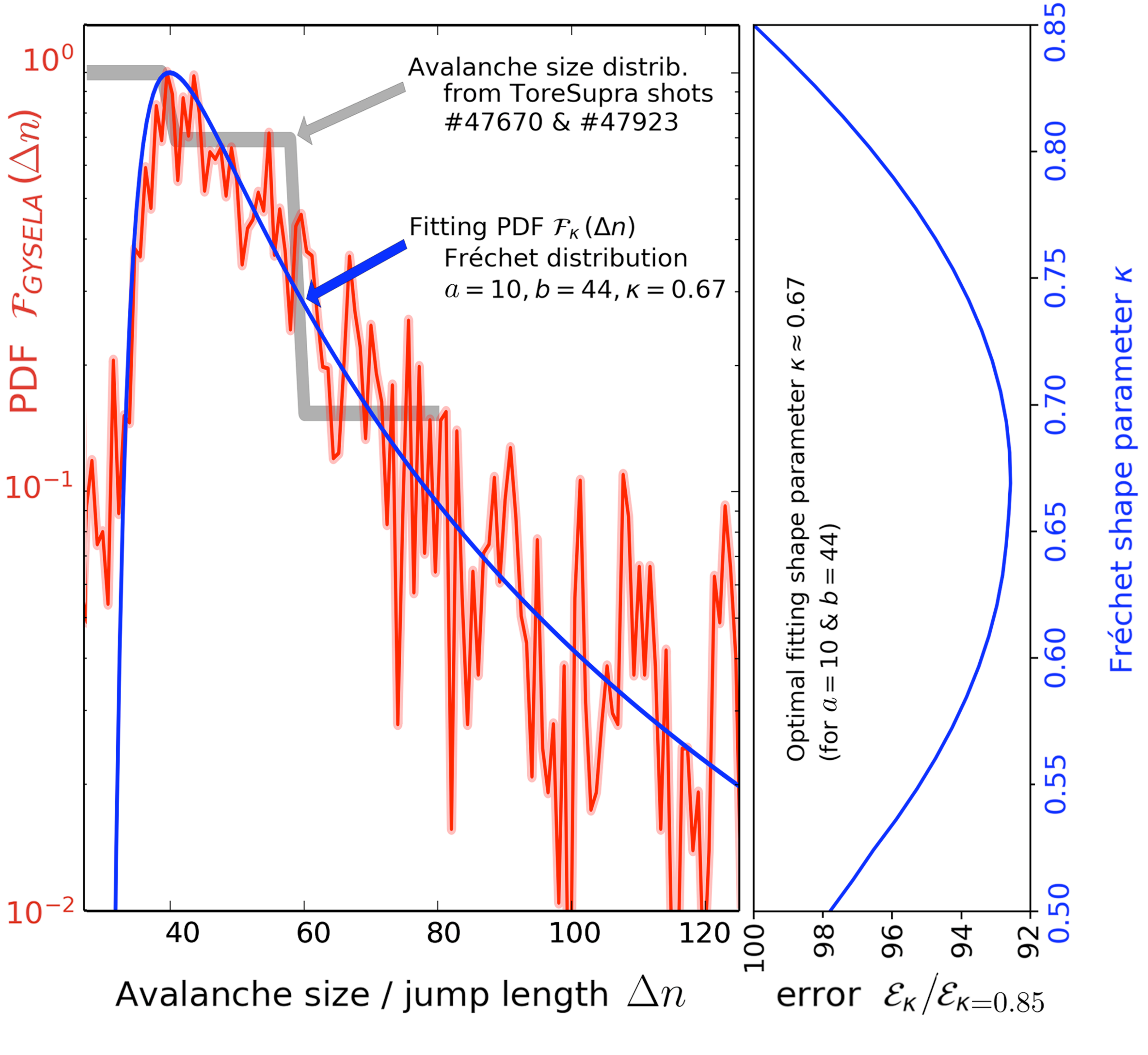}
\caption{\label{} The \textsc{Gysela} computed probability density (red color) versus the Fr\'echet distribution with $\kappa \simeq 0.67$ for which the optimum fit was obtained. Coarse experimental distribution is plotted in gray color. The right panel summarizes the normalized root-mean-square error in a percentile to the maximum error at $\kappa \simeq 0.85$, showing that the normalized error is minimized for $\kappa \simeq 0.67$. Adapted from Ref. \cite{PRE21}.     
}
\end{figure}

In the analysis of Ref. \cite{PRE21}, a somewhat different estimate on the shape parameter $\kappa$ was obtained theoretically, namely $\kappa = (\sqrt{17} + 1) / 8 \simeq 0.64$. This estimate is numerically close to, yet is analytically dissimilar from, the result $\kappa = 2/3$ suggested by the comb model. In particular, the value $\kappa = 2/3$ is a rational number, while $\kappa = (\sqrt{17} + 1) / 8$ is an irrational one.  

The discrepancy between the two values is not really surprising. In fact, the analysis of Ref. \cite{PRE21} was built on a mean-field approximation \cite{Leggett} and uses as a basis the nonlinear Schr\"odinger equation with subquadratic nonlinearity \cite{PRE19,PRE23}. In contrast, the transport model in Eq.~(\ref{KEK-SP}) assumes no reduction to mean-field properties and refers directly to the three-wave interaction Hamiltonian in Eq.~(\ref{3W}). As a consequence, the comb model offers a more precise estimate on $\kappa$ (or, at least, a better fit to simulations) by connecting the turbulence spreading dynamics to backbone transport on a Dirac comb. The end result is $\kappa = 1/\mu = 2/3$.    

\section{Fractional relaxation equation}

Performing the Fourier transform of the bi-fractional Eq.~(\ref{KEK-RL}) one obtains the fractional relaxation equation \cite{Klafter,Sokolov} 
\begin{equation}
\frac{d}{d t} \hat f (k, t) =-\tau_k^{-\beta/2}\,_{0}{D}_t^{1-\beta/2} \hat f (k, t),
\label{Relax} 
\end{equation}
where $\hat f (k, t)$ denote the Fourier components of $f (x, t)$ and we have introduced 
\begin{equation}
\tau_k^{-\beta/2} = |k|^{\mu}K_{\beta, \mu}.
\label{Tau} 
\end{equation}
In writing Eq.~(\ref{Relax}) we took into account that the Fourier transform of the Riesz fractional derivative $\partial^\mu / \partial |x|^\mu$ is $-|k|^{\mu}$, where one suppresses the imaginary unit $i^{\mu}$ following a convention used in fractional calculus (Ref. \cite{Klafter}, p. 26). 

The solution of the fractional relaxation equation~(\ref{Relax}) satisfying the initial condition $\hat f (k, t=0) = 1$ is given by the Mittag-Leffler function \cite{Klafter,Samko,Springer}
\begin{equation}
E_{\beta/2} [-(t/\tau_k)^{\beta/2}] = \sum_{m=0}^{\infty}\frac{[-(t/\tau_k)^{\beta/2}]^m}{\Gamma(1+\beta m / 2)},
\label{ML} 
\end{equation}
where $\Gamma$ denotes the Euler gamma function. For $t\gg \tau_k$, the Mittag-Leffler function $E_{\beta/2} [-(t/\tau_k)^{\beta/2}]$ is approximated by a power law 
\begin{equation}
E_{\beta/2} [-(t/\tau_k)^{\beta/2}] \simeq \frac{1}{\Gamma(1-\beta/2)}(t/\tau_k)^{-\beta/2},
\label{MLLT} 
\end{equation}
showing that $\hat f (k, t) \simeq [(t/\tau_k)^{\beta/2}\Gamma(1-\beta/2)]^{-1}$ for $t\rightarrow+\infty$. Assuming three-wave interactions ($\beta = 2/3$) one has $\hat f (k, t) \propto (t/\tau_k)^{-1/3}$, while for four-wave interactions ($\beta = 1$) one obtains $\hat f (k, t) \propto (t/\tau_k)^{-1/2}$. 

These theory findings can be supported by experimental results from the CASTOR tokamak \cite{Mart}, according to which the relaxation function has a power-law shape $\varphi_\alpha (\tau) \simeq (\tau/\tau_0)^{-\alpha}$, with the $\alpha$ value ranging between 0.3 and 0.5 depending on parameters of the plasma discharge and the time interval that is analyzed. We interpret this conformity to the CASTOR measurements as a confirmation that the relaxation dynamics is non-Markovian and involves a long-time power-law tail consistently with the Mittag-Leffler relaxation pattern in Eq.~(\ref{MLLT}). 

Another point of interest here is that the distribution of waiting times in Eq.~(\ref{WT}) can be translated into a power-law frequency distribution in accordance with
\begin{equation}
\chi_\alpha (\omega) = \chi_\alpha (\Delta t) \frac{d}{d\omega}\Delta t \propto \omega^{1+\alpha}\omega^{-2}\simeq \omega^{-(1-\alpha)}.
\label{FF} 
\end{equation} 
More explicitly, we have $\chi_\alpha (\omega) \propto \omega^{-2/3}$ for $\alpha = 1/3$ and $\chi_\alpha (\omega) \propto \omega^{-1/2}$ for $\alpha = 1/2$. Such frequency spectra have been observed in the edge region of different tokamaks \cite{Diam95,Sanchez,Carr1,Carr2,Pol99} and discussed in terms of self-organized criticality \cite{Bak1,Bak2,Jensen}.

\section{Summary and Final Remarks}

In summary, we have proposed a model of turbulence spreading driven by inelastic resonant interactions of waves on a lattice. The theory model, which we discuss, was inspired by the studies of quantum localization of dynamical chaos \cite{Sh93,PS,Flach,EPL,QW,PRE17}. Yet so, it presents a few particularities, which may be seen in both the basic configuration, which we consider, and the mathematical formalism, which we apply. 

In terms of configuration, we implement a setting according to which the noninteracting waves can propagate freely in a preferred direction, while being linearly localized in the transverse direction. A situation of this kind is accounted in tokamak plasmas where, e.g., drift wave type waves propagate mainly in the poloidal direction perpendicular to the magnetic field and are excited by localized pressure gradients in the radial direction. 

In terms of mathematical formalism, we deliberately avoided the introduction of the nonlinear Schr\"odinger equation as of Refs. \cite{PRE19,PRE21,PRE23}. Instead, we based our analysis on the Hamiltonian of inelastic wave-wave interaction and the conservation laws. This way, we could pencil the similarities and differences between three- and four-wave interaction patterns, while accounting for these patterns on essentially an equal footing. 

In terms of asymptotic transport scalings, we have seen that the asymptotic spreading is always {\it subdiffusive}, is appreciably faster in the case of three-wave interactions, and is somewhat slower in the case of four-wave interactions. The main difference is that three-wave interactions produce {\it nonlocal} dynamics with L\'evy flights, while four-wave interactions do {\it not} do so. In the latter case, the dynamics is found to be local, i.e., satisfies the defining conditions of the central limit theorem. 

More explicitly, if the interactions are three-wave-like, then the asymptotic spreading is characterized universally by a subdiffusive transport scaling $\Delta n\propto t^{1/3}$. If four-wave-like, then by the scaling $\Delta n\propto t^{1/4}$. Scaling relationships of this form hold for the number of states (in our notation, $\Delta n$). Yet, one can translate these scalings into spatial (radial) spread with the aid of $\Delta n \propto \Delta x$, and by doing so pave the way to a comparison against known scalings from numerical simulations. In the case of three-wave interactions, our model predicts $\Delta x\propto t^{1/3}$ in a remarkable agreement with the F-KPP result \cite{Zonal}. 

From an energy-budget perspective, three-wave triad interactions constitute a preferred transport channel as such interactions correspond to a lower-order correction to $H_0$, whereas four-wave interactions correspond to a higher-order correction. If both three- and four-wave interactions are allowed by the dispersion relation, then the actual spreading rate is determined by the three-wave nonlinear dynamics. 

A special case of the above theory is a situation when the interaction process involves the zero-frequency resonance between two high $\bf k$ running waves and one low-frequency standing wave. We have seen that this type of interaction process corresponds to a very peculiar dynamical pattern, in which the asymptotic dispersion is three-wave-like, while the distribution of waiting times is four-wave-like. We associated this special case of mixed statistics with the self-organization of L-mode tokamak plasma into banded flows or staircases \cite{DF2010,DF2015,DF2017,Horn2017}. 

The various transport regimes studied in this paper's work (three-wave, four-wave, mixed) are collected for comparison in Table~1, where one also finds a summary of the exponents characterizing the corresponding 1D reduced transport models. 

\begin{table}[t]
\begin{center}
\begin{tabular}{p{2.1cm}p{2.0cm}p{2.0cm}p{2.0cm}} \hline \hline
Exponent & Three-wave & Four-wave & Staircase \\ 
\hline 
$s$ & $1/2$ & 1 & Mixed\footnote{The dispersion of asymptotic transport by plasma avalanches corresponds to three-wave interactions, with $s = 1/2$, while the distribution of waiting times is such as if the interactions are four-wave, with $s=1$.} \\
$\alpha$ & $1/3$ & $1/2$ & $1/2$ \\
$H$ & $1/3$ & $1/2$ & $1/2$ \\
$\beta$ & $2/3$ & $1$ & $1$ \\
$\mu$ & $1$ & $2$ & $3/2$ \\ 
$\kappa$ & $1$ & N/A & $2/3$ \\ 
Property & Three-wave & Four-wave & Staircase \\ 
\hline
$\langle(\Delta n)^2 (t)\rangle$ & $\propto t^{2/3}$ & $\propto t^{1/2}$ & $\propto t^{2/3}$ \\
$\chi_\alpha (\Delta t)$ & $\propto (\Delta t)^{-4/3}$ & $\propto (\Delta t)^{-3/2}$ & $\propto (\Delta t)^{-3/2}$ \\
$\chi_H (\Delta t)$ & $\propto (\Delta t)^{-4/3}$ & $\propto (\Delta t)^{-3/2}$ & $\propto (\Delta t)^{-3/2}$ \\
$\chi_\mu (|\Delta x|)$ & $\propto |\Delta x|^{-2}$ & Gaussian & $\propto |\Delta x|^{-5/2}$ \\
$\chi_\mu (\ell)$\footnote{If radial transport occurs in the form of avalanches.} & $\propto \ell^{-2}$ & N/A & $\propto \ell^{-5/2}$ \\
$\chi_\alpha (\omega)$ & $\propto \omega^{-2/3}$ & $\propto \omega^{-1/2}$ & $\propto \omega^{-1/2}$ \\
$\hat f (k, t)$ & $\propto (t/\tau_k)^{-1/3}$ & $\propto (t/\tau_k)^{-1/2}$ & $\propto (t/\tau_k)^{-1/2}$ \\
Transport Eq. & Eq.~(\ref{Hil}) & Eq.~(\ref{KEK-RL+}) & Eq.~(\ref{KEK-SP}) \\
Non-Markov & Yes, $\alpha = \frac{1}{3}$ & Yes, $\alpha = \frac{1}{2}$ & Yes, $\alpha = \frac{1}{2}$ \\
Nonlocal &Yes, $\mu = 1$ & No\footnote{The general conditions of the central limit theorem apply \cite{Gnedenko,Bouchaud}.}, $\mu = 2$ & Yes, $\mu = \frac{3}{2}$ \\
L\'evy flights & Yes\footnote{Special value corresponding to Cauchy-L\'evy flights \cite{Chechkin,Rest}.}, $\mu = 1$ & No\footnote{Corresponds to the local limit of the Riesz fractional derivative and Gaussian distribution of the jump lengths \cite{Rest,Chechkin}.}, $\mu = 2$ & Yes, $\mu = \frac{3}{2}$ \\
\hline \hline
\end{tabular}
\end{center}
\caption{A summary of results and comparison between three- and four-wave interaction patterns. By examining the corresponding transport exponents one sees that spreading is faster in the case of three-wave interactions, in which case the asymptotic dynamics is nonlocal (involves L\'evy flights). The special case of mixed statistics (i.e., the zero-frequency resonance) is reported separately under the heading Staircase. We associate this special case with the self-organization of L-mode tokamak plasma into banded flows or staircases. N/A means the quantity is not well-defined in the case of four-wave interactions.} \label{tab1}
\label{default}
\end{table}

By examining Table~1 one sees that a common feature among all regimes is non-Markovianity, i.e., the nonlinear wave dynamics is long-time correlated in all cases. This is indeed characteristic and finds explanation \cite{PS,PRE23} in the simultaneous presence of domains of chaotic and regular motion, leading to some nonergodicity of asymptotic spreading \cite{PRE24}. In contrast, the nonlocal signatures are more restrictive in that they occur specifically by way of three-wave interactions or the zero-frequency resonance, but {\it not} through four-wave interactions.   

Further inspection of Table~1 suggests a set of unique signatures or fingerprints of three-wave interactions: a relatively fast asymptotic spreading complying with an $\Delta x\propto t^{1/3}$ scaling; an explicitly nonlocal behavior with Cauchy-L\'evy flights; and an algebraic, rather than exponential, tunneling pattern. The fingerprints of four-wave interactions are, on the contrary, a slower spreading conforming to an $\Delta x\propto t^{1/4}$ behavior; the absence of flights; and an exponentially decaying density of the probability to spill over a barrier. Both spreading patterns appear to be non-Markovian, with a distribution of trapping times.  

An important conclusion to be drawn from the above analysis is that both spreading and staircasing can be described based on the same mathematical formalism, using the Hamiltonian of inelastic wave-wave interaction and a mapping procedure into the comb space. From this perspective, one dares say {\it the plasma staircase is a very special case of turbulence spreading}, which is mediated by the zero-frequency resonance in Eq.~(\ref{R0}). This observation explains the observed involvement \cite{Short,Nature,Korean} of turbulence spreading in the formation of staircase dynamical patterns and coupling to transport barriers.  

From a kinetic perspective, we have seen that resonant wave-wave interactions, if three- or four-wave-like, induce a transport process of the CTRW type \cite{CTRW1,CTRW2,Bouchaud} and therefore lead to a theoretical description in terms of fractional-derivative equations \cite{Klafter,Sokolov,Rest}. In that regard, the comb approach is fundamental as it paves the way to obtaining the fractional exponents of these equations {\it exactly} by mapping the lattice dynamics onto the Dirac comb. 

Focusing on time-fractional diffusion Eq.~(\ref{KEK-RL+}), we have seen that the Riemann-Liouville derivative on the right-hand side of this equation stems from a reduction to 1D of a more general 2D Fokker-Planck equation with delta-coupling between the degrees of freedom. The observed connection between 1D equation with fractional differentiation over time and a 2D Fokker-Planck equation with coupling sheds light on theoretical foundations of fractional kinetics \cite{Klafter,Sokolov}, an important topic in statistical physics of complex systems \cite{Rest,Andrey,IomBas}.  

By applying combs to the staircase transport problem \cite{DF2010,DF2015}, we inferred an estimate on the shape parameter $\kappa$, which characterizes the observed distributions \cite{DF2017,Horn2017,PRE21} of plasma avalanches over their sizes (in terms of the Fr\'echet distribution \cite{Haan}). Our results indicate that the $\kappa$ value pertaining to these distributions is given by the inverse L\'evy index $\mu$, i.e., $\kappa = 1/\mu$. Because, by its definition \cite{Bouchaud,Klafter}, $\mu < 2$, we have $\kappa > 1/2$, provided the distribution of the jump lengths in Eq.~(\ref{PLF}) is L\'evy stable, as it should \cite{Gnedenko}.  

More so, we have discussed that staircasing\textemdash i.e., the emergence of quasiregular banded flows \cite{Zonal,DF2010,DF2015,McIntyre} from a micro-turbulence background\textemdash is driven universally by the zero-frequency resonance in Eq.~(\ref{R0}). In that regard, the interaction Hamiltonian~(\ref{3W+a}) with intermediate bound states predicts that the asymptotic spreading dynamics is nonlocal, with L\'evy flights, and is characterized by the very specific value of the L\'evy index $\mu = 3/2$. Using the general \cite{PRE18} relation $\kappa = 1/\mu$ one gets $\kappa = 2/3$. This theoretical result granted by the comb approximation meets gracefully the numerical estimate $\kappa \simeq 0.67$ \cite{PRE21} obtained from flux-driven gyrokinetics.   

Overall, we have seen that the subdiffusive transport scaling $\langle(\Delta x)^2(t)\rangle \propto t^{2/3}$ is an important lower bound on radial transport across the plasma staircase \cite{PRE21}. We associate this lower bound with the confining effect of coupled transport barriers in vicinity of marginality.  

Finally, we remark that the comb model\textemdash by its very construction\textemdash proves to be an efficient and simple way to characterize the anisotropic transport processes in electrostatic drift-wave turbulence. Indirectly, it also characterizes the anisotropic particle dispersion in beta-plane turbulence due to the similarity between drift waves in plasmas and Rossby waves on the beta-plane \cite{Rossby}. In that respect, we note that such anisotropic transport has been studied numerically for electrostatic drift waves in Refs. \cite{JJR,PLAN,Basu1,Basu2} and for fluid turbulence in Refs. \cite{Babi,Ann}. 

\begin{acknowledgments}
Helpful discussions with P.~H.~Diamond, G.~Dif-Pradalier, X.~Garbet, and Y.~Sarazin are acknowledged with thanks. One of us (A.V.M.) would like to thank the Isaac Newton Institute for Mathematical Sciences, Cambridge, U.K., for support and hospitality during the programmes ``Anti-diffusive dynamics: from sub-cellular to astrophysical scales" and ``Stochastic systems for anomalous diffusion," where work on this paper was undertaken. This work was supported by EPSRC grants No. EP/R014604/1 and EP/Z000580/1. Partial support was received from a grant from the Simons Foundation. 
\end{acknowledgments}



\appendix*

\section{The Fokker-Planck model with singular backbone term} 

Let us rewrite the Fokker-Planck equation~(\ref{KEK+N}) first in a more suitable for our analysis form 
\begin{equation}\label{A1}
\frac{\partial}{\partial t} f(x,y,t) = \left[\varrho\frac{\partial^2}{\partial y^2} + \delta(y)\frac{\partial^2}{\partial x^2}\right] f(x,y,t),
\end{equation}
where $\varrho \propto K_1/ K_2$ is the normalized coefficient of the transport process, and $t$ is the dimensionless time. Focusing on the backbone transport, because the Fokker-Planck equation in Eq.~(\ref{A1}) is symmetric with respect to the inversion $x\rightarrow -x$, it is convenient to set the coordinate $x$ on the semi-axis $0 \leq x < + \infty$. This setting is technical and does not influence results. Concerning the coordinate $y$, we do not assume any restriction, thus letting $-\infty < y < +\infty$. As the random walk starts at the origin $(x=0, y=0)$, we may also impose 
\begin{equation}
f(x=+\infty,y,t)=\partial_x f(x=+\infty,y,t)=0,
\label{BC1} 
\end{equation} 
\begin{equation}
f(x, y=\pm\infty,t)=\partial_y f(x,y=\pm\infty)=0,
\label{BC2} 
\end{equation} 
where $\partial_x$ and $\partial_y$ denote the partial derivatives along $x$ and $y$, respectively. Concerning the boundary condition at $x=0$, it is assumed that $f(x=0, y=0, t) = C_0$, where $C_0={\rm const}$ for all $t > 0$. The initial condition for $t=0$ is defined as $f_0 (x, y) \equiv f(x, y, t=0) = C(x)\delta(y)$, where $C(x)=0$ for $x > 0$. 

The transport problem is solved straightforwardly by applying the Laplace transform with respect to time to both sides of the Fokker-Planck equation~(\ref{A1}), yielding 
\begin{equation}\label{hpcm-1} 
s\hat{f}(x,y,s) = \left[\varrho\frac{\partial^2}{\partial y^2} + \delta(y)\frac{\partial^2}{\partial x^2}\right] \hat{f}(x,y,s),
\end{equation}
where $\hat{f}(x,y,s)=\hat\mathcal{L}\left[f(x,y,t)\right](s)$ is the Laplace image of $f(x,y,t)$. Equation~(\ref{hpcm-1}) is supplied with the Laplace image of the boundary condition at $x=0$, i.e.,   
\begin{equation}\label{hpcm-2} 
\hat{f}(x=0,y=0,s) =\int_0^{+\infty}e^{-st} f(x=0,y=0,t) dt,
\end{equation}
from which $\hat{f}(x=0,y=0,s)= {C_0}/{s}$. The solution to Eq.~(\ref{hpcm-1}) is obtained using the Ansatz 
\begin{equation}\label{hpcm-3} 
\hat{f}(x,y,s) = e^{-|y|\sqrt{s/\varrho}} \phi (x,s), 
\end{equation}
where $\phi (x,s)$ has the sense of a probability distribution along the backbone in the Laplace space. Setting $y = 0$ in Eq.~(\ref{hpcm-3}), one infers $\phi (x,s) = \hat{f}(x,y=0,s)$, where $\hat{f}(x,y=0,s) = \hat\mathcal{L}\left[f(x,y=0,t)\right](s)$ is the Laplace image of $f(x,y=0,t)$. Mathematically, it is convenient to introduce $f_1(x,t)=\int_{-\infty}^{+\infty}f(x,y,t) dy$, which integrates the probability density $f(x,y,t)$ in side branches. In the Laplace domain, the function $f_1(x,t)$ becomes
\begin{equation}\label{hpcm-6} 
\hat{f}_1(x,s) = \hat{f}(x,y=0,s) \int_{-\infty}^{+\infty}e^{-|y|\sqrt{s/\varrho}}dy, 
\end{equation}
leading to
\begin{equation}\label{hpcm-6+} 
\hat{f}_1(x,s) = 2\left[{\varrho}/{s}\right]^{{1}/{2}}\phi(x,s), 
\end{equation}
where Eq.~(\ref{hpcm-3}) has been considered. Integrating both sides of Eq.~(\ref{hpcm-1}) over $y$ in infinite limits from $-\infty$ to $+\infty$, and applying the boundary conditions in Eqs.~(\ref{BC1}) and~(\ref{BC2}), one gets
\begin{equation}\label{hpcm-5a} 
s\hat{f}_1(x,s)= \frac{\partial^2}{\partial x^2} \hat{f}(x,y=0,s),
\end{equation}
from which, on account of Eq.~(\ref{hpcm-3}), 
\begin{equation}\label{hpcm-5a} 
s\hat{f}_1(x,s)= \frac{\partial^2}{\partial x^2} \phi (x,s)
\end{equation}
and
\begin{equation}\label{hpcm-5b} 
\phi(x=0,s)=\hat{f}(x=0,y=0,s) =\frac{C_0}{s}. 
\end{equation}
Combining Eqs.~(\ref{hpcm-5a}) and~(\ref{hpcm-6+}), and solving the ensuing differential equation for $\phi (x, s)$, one gets, with the aid of Eq.~(\ref{hpcm-5b}), 
\begin{equation}\label{hpcm-7} 
\phi(x,s)=\frac{C_0}{s}\exp\left[-(4s\varrho)^{1/4}x\right].
\end{equation}
The mean-squared displacement along the backbone is obtained by taking the second moment of the probability density function $\phi(x,t)$, i.e., 
\begin{equation}\label{hpcm-9} 
\langle (\Delta x)^2(t)\rangle=\frac{1}{N(t)}\int_0^{+\infty}x^2 \phi (x,t)dx,
\end{equation}
where $\phi (x, t) = \hat\mathcal{L}^{-1}\phi(x, s)$, $\phi(x,s)$ is given by Eq.~(\ref{hpcm-7}), and we have introduced 
\begin{equation}\label{hpcm-8} 
N (t) = \int_0^{+\infty}\phi(x,t)dx = \hat\mathcal{L}^{-1}\left[\int_0^{+\infty}\phi(x,s)dx\right]
\end{equation}
to normalize $\phi (x,t)$. A simple calculation leads to 
\begin{equation}\label{hpcm-88} 
N (t) = C_0\hat\mathcal{L}^{-1}\left[\frac{1}{(4\varrho)^{1/4}}s^{-5/4}\right].
\end{equation}
From Eq.~(\ref{hpcm-9}) one also gets 
\begin{equation}\label{hpcm-99} 
\langle (\Delta x)^2(t)\rangle=\frac{1}{N(t)}\hat\mathcal{L}^{-1}\int_0^{+\infty}x^2 \phi (x,s)dx.
\end{equation}
Substituting $\phi (x,s)$ from Eq.~(\ref{hpcm-7}), and performing the improper integration over $x$, one obtains
\begin{equation}\label{hpcm-999} 
\langle (\Delta x)^2(t)\rangle=2C_0\frac{1}{N(t)} \hat\mathcal{L}^{-1}\left[\frac{1}{(4\varrho)^{3/4}}s^{-7/4}\right].
\end{equation}
Calculating the inverse Laplace transform in Eqs.~(\ref{hpcm-88}) and~(\ref{hpcm-999}), one finds  
\begin{equation}\label{hpcm-9ff} 
N(t) \simeq (C_0 / \sqrt{2}) (t/\varrho)^{1/4},
\end{equation}
\begin{equation}\label{hpcm-9ff+} 
\langle (\Delta x)^2(t)\rangle \simeq \frac{C_0 / \sqrt{2}}{N(t)} (t/\varrho)^{3/4},
\end{equation}
from which, by combining Eqs.~(\ref{hpcm-9ff}) and~(\ref{hpcm-9ff+}),
\begin{equation}\label{hpcm-9f} 
\langle (\Delta x)^2(t)\rangle \simeq A_\varrho (t/\varrho)^{1/2},
\end{equation}
where $A_\varrho$ is a numerical coefficient of the order of 1. 

To reconstruct the full (two-dimensional) probability density in the comb space, i.e., the $f = f(x,y,t)$ function, one needs to tame
\begin{equation}\label{hpcm-10}
f(x,y,t)=C_0\mathcal{L}^{-1}\left[\frac{1}{s}e^{-|y|\sqrt{s/\varrho}} e^{-x(4s\varrho)^{1/4}}\right]
\end{equation}
with the new normalization
\begin{equation}\label{hpcm-111}
N^\prime(t)=C_0\mathcal{L}^{-1}\left[\frac{1}{s}\int_0^{+\infty}e^{-x(4s\varrho)^{1/4}}dx\int_{-\infty}^{+\infty} e^{-|y|\sqrt{s/\varrho}}dy\right].
\end{equation}
Straightforward integration yields
\begin{equation}\label{hpcm-11}
N^\prime(t)={\sqrt{2}}C_0\mathcal{L}^{-1}\left[\varrho^{1/4}s^{-7/4}\right]. 
\end{equation} 
Upon Laplace inversion to the time domain,
\begin{equation}\label{hpcm-11+}
N^\prime(t) \simeq {\sqrt{2}}C_0\varrho (t / \varrho)^{3/4},
\end{equation} 
where we omitted for simplicity a numerical coefficient due to the Euler gamma function. The mean-squared displacement in the $x$ direction is written as the double integral
\begin{equation}\label{hpcm-12}
\langle (\Delta x)^2(t)\rangle=\frac{1}{N^\prime(t)}\mathcal{L}^{-1}\left[
\int_0^{+\infty}\int_{-\infty}^{+\infty}x^2 \hat f(x,y,s)dxdy\right].
\end{equation}
The expression is square brackets is expanded as 
\begin{equation}\label{hpcm-12+}
\Big[\dots\Big]=
\frac{C_0}{s}
\int_0^{+\infty}x^2e^{-x(4s\varrho)^{1/4}}dx\int_{-\infty}^{+\infty} e^{-|y|\sqrt{s/\varrho}}dy,
\end{equation}
where Eqs.~(\ref{hpcm-3}) and~(\ref{hpcm-7}) have been considered. Integrating over $x$ and $y$ in infinite limits, after a simple algebra one obtains
\begin{equation}\label{hpcm-12++}
\langle (\Delta x)^2(t)\rangle={\sqrt{2}}C_0\frac{1}{N^\prime(t)}\mathcal{L}^{-1}\left[
\varrho^{-1/4}{s}^{-9/4}\right].
\end{equation}
Calculating the inverse Laplace transform, one gets
\begin{equation}\label{hpcm-12+L}
\langle (\Delta x)^2(t)\rangle \simeq {\sqrt{2}}C_0\varrho\frac{1}{N^\prime(t)} (t / \varrho)^{5/4}.
\end{equation}
Now substituting $N^\prime (t)$ from Eq.~(\ref{hpcm-11+}), one recovers the asymptotic scaling law 
\begin{equation}\label{hpcm-12+AS}
\langle (\Delta x)^2(t)\rangle \simeq A_\varrho (t / \varrho)^{1/2}
\end{equation}
consistently with Eq.~(\ref{hpcm-9f}). $A_\varrho$ is obtained by keeping everywhere the coefficients dictated by Euler's integral and is left as an exercise to the reader.  

One sees that the Fokker-Planck equation~(\ref{A1}) leads to the same subdiffusive transport scaling $\langle (\Delta x)^2(t)\rangle \propto t^{1/2}$ as the fractional kinetic Eq.~(\ref{KEK-RL+}) does. To this end, the use of fractional-derivative equations becomes a matter of mathematical taste: The reduction to 1D of the original 2D transport model~(\ref{KEK+N}) comes at a cost of introducing the Rimeann-Liouville derivative~(\ref{R-L}) and recognizing the elegance of the Fox $H$-function in Eq.~(\ref{Fox}).


\nocite{*}

\end{document}